\documentclass[useAMS,usenatbib]{mn2e}

\usepackage{epsfig}
\usepackage{deluxetable}

\def\h2{H$_2$}
\def\f0{$F_0$}

\newcommand\ion[2]{#1$\;${\small\rmfamily\@Roman{#2}}\relax}%

\title[GRB\,120404A Jet Geometry and Microphysics ]{New constraints on GRB jet geometry and relativistic shock physics}
\author[C.~Guidorzi et~al.]{C.~Guidorzi$^{1}$\thanks{E-mail:guidorzi@fe.infn.it},
  C.~G.~Mundell,$^{2}$ 
 R.~Harrison,$^{2}$ R.~Margutti,$^{3}$ V.~Sudilovsky,$^{4}$ 
\newauthor B.~A.~Zauderer,$^{3}$ S.~Kobayashi,$^{2}$ A.~Cucchiara,$^{5}$ A.~Melandri,$^{6}$ S.~B.~Pandey,$^{7}$
\newauthor   E.~Berger,$^{3}$ D.~Bersier,$^{2}$ V.~D'Elia,$^{8}$ A.~Gomboc,$^{9,10}$ J.~Greiner,$^{4}$ J.~Japelj,$^{9}$
\newauthor  D.~Kopa{\v c},$^{9}$ B.~Kumar,$^{7}$ D.~Malesani,$^{11}$ C.~J.~Mottram, $^{2}$ P.~T.~O'Brien,$^{12}$
\newauthor   A.~Rau,$^{4}$ R.~J.~Smith,$^{2}$ I.~A.~Steele,$^{2}$ N.~R.~Tanvir,$^{12}$ F.~Virgili$^{2}$\\
\mbox{}\\
$^{1}$Department of Physics and Earth Sciences, University of Ferrara, via Saragat 1,
  I-44122, Ferrara, Italy\\
$^{2}$Astrophysics Research Institute, Liverpool John Moores University, IC2, Liverpool Science Park,
      146 Brownlow Hill, Liverpool L3 5RF, UK\\
$^{3}$Harvard--Smithsonian Center for Astrophysics, 60 Garden Street, Cambridge,
  MA 02138, USA\\
$^{4}$Max--Planck--Institut f\"ur extraterrestrische Physik, Giessenbachstrasse 1, I-85748, Germany\\
$^{5}$Department of Astronomy and Astrophysics, UCO/Lick Observatory, University of California, 1156 High Street,
Santa Cruz, CA 95064, USA\\
$^{6}$INAF -- Osservatorio Astronomico di Brera, via E. Bianchi 46,
  I-23807 Merate (LC), Italy\\
$^{7}$Aryabhatta Research Institute of Observational Sciences, Manora Peak, Nainital, Uttarakhand, India, 263129\\
$^{8}$ASI Science Data Center, via Galileo Galilei, I-00044, Frascati, Italy\\
$^{9}$Faculty of Mathematics and Physics, University of Ljubljana,
  Jadranska 19, SI-1000 Ljubljana, Slovenia\\
$^{10}$Centre of Excellence SPACE-SI, A\v sker\v ceva cesta 12, SI-1000
  Ljubljana, Slovenia\\
$^{11}$Dark Cosmology Centre, Niels Bohr Institute, University of Copenhagen,
  Juliane Maries vej 30, DK-2100 K\o behavn \O , Denmark\\
$^{12}$Department of Physics and Astronomy, University of Leicester,
  University Road, Leicester LE1 7RH, UK\\
}

\begin{document}

\date{\today}


\maketitle

\label{firstpage}

\begin{abstract}

We use high--quality, multi-band observations of {\em Swift} GRB\,120404A,
from $\gamma$-ray to radio frequencies, together with the new hydrodynamics
code of van~Eerten et al. (2012) to test the standard synchrotron shock model.
The evolution of the radio and optical afterglow,
with its prominent optical rebrightening at t$_{\rm rest}\sim 260$--$2600$~s,
is remarkably well modelled by a decelerating jet viewed close to the jet edge,
combined with some early re--energization of the shock.
We thus constrain the geometry of the jet with half--opening
and viewing angles of $23^\circ$ and $21^\circ$ respectively and suggest that wide
jets viewed off-axis are more common in GRBs than previously thought.
We also derive the fireball microphysics parameters $\epsilon_B=2.4\times10^{-4}$
and $\epsilon_e=9.3\times10^{-2}$ and a circumburst density of $n=240$~cm$^{-3}$.
The ability to self--consistently model the microphysics parameters and
jet geometry in this way offers an alternative to trying to identify elusive
canonical jet breaks at late times. 
The mismatch between the observed and model-predicted X--ray fluxes is explained
by the {\em local} rather than the global cooling approximation in the synchrotron
radiation model, constraining the microphysics of particle acceleration taking
place in a relativistic shock and, in turn, emphasising the need for a more
realistic treatment of cooling in future developments of theoretical models.
Finally, our interpretation of the optical peak as due to the
passage of the forward shock synchrotron frequency highlights the importance of
high quality multi--band data to prevent some optical peaks from being
erroneously attributed to the onset of fireball deceleration. 
\end{abstract}

\begin{keywords}
gamma-rays: bursts --- radiation mechanisms: nonthermal
\end{keywords}

\section{Introduction}
\label{sec:intro}
The observational picture of the gamma--ray burst (GRB) phenomenon has constantly
been evolving during the last fifteen years since the discovery of the
long--lived afterglow radiation in the aftermath of the prompt high--energy
emission (see \citealt{Gehrels12} for a recent review).
The knowledge of the GRB host galaxies as well as of the circumburst environment
properties has been providing important clues to characterise the stellar progenitors,
to identify key factors such as metallicity (e.g., see \citealt{Fynbo12,Savaglio12}
for recent reviews), especially whenever a possible associated supernova component cannot
be observed due to distance constraints.

In the {\em Swift} and {\em Fermi} era, the phenomenology displayed across
the electromagnetic spectrum by GRB afterglows appears to be more complex
than predicted in the pre--{\em Swift} epoch \citep{Melandri08}.
In particular, clear--cut achromatic breaks in the light curves associated with the jet
angle have turned out to be unexpectedly rare events \citep{Racusin09}.
Likewise, the unexpected paucity of early optical light curves with evidence
for reverse shock (RS) emission \citep{Roming06} raised the issue of the magnetic energy
density entrained in the ejecta as a possible explanation \citep{ZK05} in addition to other
alternatives (e.g., \citealt{Mundell07,JF07,Melandri10,Guidorzi11b}).

Although in many cases fitting full data sets into a self--consistent
description of the afterglow evolution proved very problematic
(e.g., \citealt{Covino10,Gendre10}), overall the afterglow emission can be accounted
for as synchrotron with possible Inverse Compton contributions by the electrons shocked
by the GRB blast wave (e.g., \citealt{Meszaros06}), with occasional energy injection
(e.g., \citealt{Rossi11}) and/or the combination of geometric effects
(e.g., \citealt{Guidorzi09,Kruehler09,Margutti10}).

In this paper we provide a self--consistent picture of the broadband data set
we collected on GRB\,120404A, spanning from radio to X--rays, within the first few days
after the GRB itself. To this aim, we fitted the entire data set using the hydrodynamical
code recently developed by \citet{Vaneerten12b}, which models the synchrotron emission
from a relativistic fireball sweeping up homogeneous interstellar medium (ISM) within
a uniform conical structure jet with sharp edge. The excellent quality of our data
set, combined with the observed complex behaviour, represents a rigorous test
for the model and offers the opportunity to strictly constrain the energetics,
the geometry of the jet, and the microphysics parameters of the shocks.
This is one of the first cases in which a realistic (i.e. based on realistic
hydrodynamical simulations and not purely analytical)
model is applied to a broadband high--quality data set of a GRB.
We also present spectroscopic data of the optical afterglow which allowed us
to measure its redshift.

Throughout the paper, times are given relative to the GRB
trigger time of {\it Swift}/BAT, which corresponds to April 4,
2012, 12:51:02~UT. The convention $F(\nu,t) \propto
\nu^{-\beta}\,t^{-\alpha}$ is followed, where $F$ is the flux density,
the energy index $\beta$ is related to the photon index by $\Gamma=\beta + 1$.
We adopted the standard cosmological model: $H_0=71$\,km\,s$^{-1}$\,Mpc$^{-1}$,
$\Omega_\Lambda=0.73$, $\Omega_{\rm M}=0.27$ \citep{Spergel03}.

All of the quoted errors are given at 90\% confidence level for one
interesting parameter ($\Delta \chi^2 = 2.706$), unless stated
otherwise.

\section{Observations}
\label{sec:obs}

GRB\,120404A was detected and localised in real time with the {\it Swift}
Burst Alert Telescope (BAT; \citealt{Barthelmy05}) instrument \citep{Stratta12}
with an accuracy of $3\arcmin$.  The $\gamma$-ray prompt emission in the 15--150~keV
energy band lasted about 50~s.  A quick-look analysis gave a peak flux of
$(1.2\pm0.2)$~ph~cm$^{-2}$~s$^{-1}$, a fluence of about $10^{-6}$~erg~cm$^{-2}$,
and burst coordinates $\alpha$(J2000) $= 15^{\rm h} 40^{\rm m}
00\fs4$, $\delta$(J2000) $= +12^\circ 52\arcmin 57\arcsec$ with an
error radius of $1.2\arcmin$ \citep{Ukwatta12}.

The {\it Swift} X-Ray Telescope (XRT; \citealt{Burrows05}) began observing
at 130~s and promptly found a bright, uncatalogued X-ray source within the
BAT error circle.
The X-ray source position was later refined using the XRT--UVOT alignment
and matching UVOT field sources to the USNO-B1 catalogue, with burst
coordinates $\alpha$(J2000) $= 15^{\rm h} 40^{\rm m}
02\fs28$, $\delta$(J2000) $= +12^\circ 53\arcmin 04\farcs9$ with an
error radius of $1\farcs6$ \citep{Osborne12}.

The {\it Swift} Ultraviolet/Optical Telescope (UVOT; \citealt{Roming05})
began observing at 139~s and from a 147-s exposure in the white filter
found an optical candidate with magnitude $19.43\pm0.12$ with coordinates
$\alpha$(J2000) $= 15^{\rm h} 40^{\rm m} 02\fs29$, $\delta$(J2000)
$= +12^\circ 53\arcmin 06\farcs3$ with an
error radius of $0\farcs65$ \citep{Stratta12,Breeveld12}.

The UVOT optical candidate was soon confirmed independently by
the Faulkes Telescope North (FTN), which began observing 4~minutes
after the GRB trigger time \citep{Guidorzi12b}.
We measured the redshift of $z=2.876$ with Gemini-North about 1~hour
after the burst upon the identification of several absorption lines
\citep{Cucchiara12}; this value was later confirmed with the X--shooter
instrument \citep{Xshooter} at the ESO Very Large Telescope (VLT),
which observed at 16~hours post burst \citep{DElia12}.

The optical afterglow was observed by a number of facilities: the
FTN and the Faulkes Telescope South (FTS) jointly monitored it from
4~minutes to $5.5$~hours with the $BVRi'$ filters. The optical light curve
exhibited a rebrightening which peaked around 40~minutes post burst with
a magnitude of $R=16.9$, as also noted by others \citep{Tristram12}.

We kept monitoring the afterglow with the Gamma-Ray Burst Optical and
Near-Infrared Detector (GROND; \citealt{Greiner08}) which started
simultaneous observations in $g'r'i'z'JHK$ filters at $18.2$~hours
after the burst \citep{Sudilovsky12}. We also collected data with the $1.04$-m telescope
at the Aryabhatta Research Institute of observational sciences (ARIES)
in Nainital, India, starting from $6.5$ hours post GRB with $RI$ filters \citep{Kumar12}.

Finally, we discovered the radio counterpart with the Karl G. Jansky Very Large Array
(VLA; \citealt{PerleyEVLA}) at 22~GHz at $0.75$~days at the position
$\alpha$(J2000) $= 15^{\rm h} 40^{\rm m} 02\fs28$ ($\pm 0.01$),
$\delta$(J2000) $= +12^\circ 53\arcmin 06\farcs1$
($\pm 0.1$) with a flux density of $(88\pm24)~\mu$Jy \citep{Zauderer12}.

The Galactic reddening along the direction to the GRB is $E_{B-V} =
0.050$~mag \citep{Schlegel98}. The Galactic extinction in each filter
has been estimated through the NASA/IPAC Extragalactic Database
extinction
calculator\footnote{\texttt{http://nedwww.ipac.caltech.edu/forms/calculator.html}.}.
Specifically, the extinction in each filter is derived through the
parametrisation by \citet{Cardelli89}: $A_{U}=0.27$, $A_B=0.22$,
$A_{g'}=0.20$, $A_V=0.16$, $A_{r'}=0.15$, $A_R=0.13$, $A_{i'}=0.11$,
$A_I=0.10$, $A_J=0.04$, $A_H=0.03$, and $A_K=0.02$~mag.

\section{Data reduction and analysis}
\label{sec:an}

\subsection{Gamma-ray data}
\label{sec:gamma}

\begin{figure}
\centering
\includegraphics[width=6cm,angle=270]{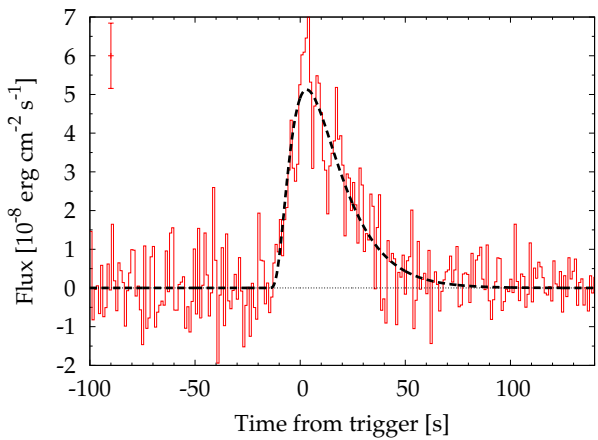}
\caption{{\it Swift}/BAT light curve in the 15--150~keV energy band
  The dashed line shows the best-fitting model as obtained with the
  model by \citet{Norris05}. The typical error bar is shown in the
upper left corner.}
\label{f:lc_gamma}
\end{figure}

We processed the {\it Swift}/BAT data of GRB\,120404A using the latest
version of the {\sc heasoft} package (v$6.12$). We extracted the mask-tagged
light curve and energy spectra in the 15--150 keV energy band by adopting
the ground refined coordinates provided by the BAT team \citep{Ukwatta12}.
The BAT detector quality map was obtained by processing the closest-in-time
enable/disable detector map. Energy calibration was applied using the
closest-in-time gain/offset file with the tool {\sc batmaskwtevt}.
Figure~\ref{f:lc_gamma} shows the mask-weighted 15--150~keV time profile
of GRB\,120404A recorded by the {\it Swift}/BAT detector. 
It consists of a single fast-rise exponential-decay (FRED) pulse peaking
at 3~s with a $T_{90}=48\pm 16$~s, from $-15.5$ to $32.5$~s.
The flux shown is derived assuming the rate-to-flux conversion obtained
from the time-integrated spectrum over the $T_{90}$ interval (see below).
Fitting the time profile with the model by \citet{Norris05} gives a satisfactory result
($\chi^2/{\rm  dof}=171/143$), as shown by the dashed line in Figure~\ref{f:lc_gamma}.
The parameters used are the peak time $t_{\rm peak}$, the peak flux $A$,
the rise and decay times $\tau_{\rm r}$ and $\tau_{\rm d}$, the pulse
width $w$, and the asymmetry $k$.
Their best-fitting values are reported in Table~\ref{tab:N05}.
The shape of the pulse with a corresponding decay-to-rise ratio of $2.2$
is very typical of classical {\em Fast Rise Exponential Decays} (FREDs) \citep{Norris96}.
%
%

The 15--150 keV peak flux is calculated from the spectrum integrated
around the peak, from $2.2$ to $4.3$~s; this can be fitted with a
power-law with a photon index of $1.9\pm0.3$ ($\chi^2/{\rm dof}=6.5/8$)
with a peak flux of $(7.3\pm1.2)\times10^{-8}$~erg~cm$^{-2}$~s$^{-1}$
and a peak photon flux of $(1.1\pm0.2)$~ph~cm$^{-2}$~s$^{-1}$.
The 15--150 keV time-integrated spectrum over the $T_{90}$ interval can
be fitted with a power law with a photon index of $\Gamma_\gamma=1.91\pm0.15$
and a total fluence of $(1.63\pm0.14)\times10^{-6}$~erg~cm$^{-2}$.
Compared with the catalogue of BAT \citep{Sakamoto11}, GRB\,120404A
is a medium burst in terms of both peak flux and fluence.
The typical low-energy and high-energy photon indices of GRB prompt
emission spectra (e.g., \citealt{Kaneko06,Sakamoto11}) suggest that
the peak energy, $E_{\rm p}$, is likely to lie either within or below
the 15--150~keV energy band.

\begin{table*}
\centering
\caption{Best-fitting parameters of the time profile of the
prompt $\gamma$-ray pulse as seen in the 15--150~keV band.}
\label{tab:N05}
\begin{tabular}{ccccccc}
\hline
$t_{\rm peak}$   & Peak flux & $\tau_{\rm r}$  & $\tau_{\rm d}$  & $w$   & $k$ & $\chi^2/$dof\\
      (s)       & ($10^{-8}$~erg~cm$^{-2}$~s$^{-1}$)      &    (s)         &      (s)      &  (s)  & &\\
\hline
$3.1\pm7.1$ & $5.1\pm0.3$ & $11.0\pm2.6$ & $24.3\pm4.2$ & $35.2\pm6.6$ & $0.38\pm0.05$ & $171/143$\\
\hline
\end{tabular}
\end{table*}

Despite the unknown value of $E_{\rm p}$, we can constrain it
from the photon index measured with BAT through the relation by
\citet{Sakamoto09} taking into account its large dispersion. 
We infer that $E_{\rm p}$ is likely to range between $\sim1$ and
$100$~keV, corresponding to an intrinsic (i.e., source-rest frame)
value for $E_{\rm p,i}$ between $\sim4$ and $\sim400$~keV.
We can make a further step by constraining the isotropic-equivalent
radiated energy $E_{\gamma,{\rm iso}}$ in the GRB rest-frame 1--$10^4$~keV energy band.
We assume the spectrum to be described with the Band function with
typical values for the photon indices, $\alpha_{\rm B}=-1$ and
$\beta_{\rm B}=-2.3$ \citep{Kaneko06}.
By propagating the uncertainty on $E_{\rm p}$ in calculating the
corresponding $E_{\gamma,{\rm iso}}$, we end up with an estimate of
$E_{\gamma,{\rm iso}}=(9\pm4)\times10^{52}$~erg.

We calculated a 3$\sigma$ upper limit to the average 15--150~keV
flux in the time interval from 200 to 800~s. This is roughly simultaneous
with a constant optical flux phase preceding the major rebrightening
and, as such, is useful to constrain the spectral index of a possible
long--lived, low--level prompt emission with an optical counterpart.
We obtained an upper limit on the average flux of
$6.8\times10^{-10}$~erg~cm$^{-2}$~s$^{-1}$, or, equivalently, 
$f_{\nu,\gamma}<2.2$~$\mu$Jy at $\nu_\gamma=1.2\times10^{19}$~Hz (50~keV).
Combined with the dust--corrected value for the $R$ band measured
during the early constant phase, $f_{\nu,R}=270$~$\mu$Jy
at $\nu_R=4.7\times10^{14}$~Hz, it turns into a lower limit
to the average optical--to--$\gamma$ spectral index,
$\beta_{{\rm opt}-\gamma}>0.5$. The observed $\beta_{{\rm opt}-\gamma}$
distribution for a large number of GRBs is consistent with
values larger than $0.5$ \citep{Yost07a,Kopac13},
unless one considers GRBs classified as dark \citep{Yost07b}, most
of which are dust--extinguished \citep{Perley09,Fynbo09,Greiner11,Zauderer13}.
Hence, the possibility of a long--lasting $\gamma$--ray emission
below the BAT sensitivity sharing a common origin with the early ($t<800$~s)
optical detection is not at odds with what is observed for most
unextinguished GRBs with measured optical and high--energy prompt emission.

\subsection{X-ray data}
\label{sec:xray}

The {\it Swift}/XRT began observing GRB\,120404A on 2012 April 04 at
12:53:25~UT, about 143~s after the trigger, and ended on 2012 April 07 at
22:39:57, with total net exposures of $117$~s in window timing (WT)
and $26.2$~ks in photon counting (PC) modes spread over 6.9 days.
The XRT data were processed following the procedure described in
\citet{Margutti13}, applying calibration and standard filtering and
screening criteria. The XRT analysis was performed in the 0.3--10~keV
energy band.

We extracted the 0.3--10~keV energy spectrum in the time interval from
10.4 to 21.1~ks; later observations did not allow us to collect enough
photons to ensure the extraction of another meaningful spectrum.
Source and background spectra were extracted from the same regions as
those used for the light curve.  Spectral channels were grouped so as
to have at least 20 counts per bin. The ancillary response files were
generated using the task {\sc xrtmkarf}. Spectral fitting was
performed with {\sc xspec} (v. 12.5).  The spectrum can be modelled
with an absorbed power law with the combination of {\sc xspec} models
{\sc wabs zwabs pow}, based on the photoelectric cross section by
\citet{Morrison83}. The Galactic neutral hydrogen column density
along the GRB direction was fixed to the value determined from 21~cm
line radio surveys: $N_{\rm HI}^{\rm Gal} = 3.4\times
10^{20}$~cm$^{-2}$ \citep{Kalberla05}.  The additional X-ray
absorption, modelled in the GRB rest frame, was found to be
$N_{{\rm HI},z} = 6.3_{-5.4}^{+6.4} \times 10^{21}$~cm$^{-2}$, very typical
of X-ray afterglow spectra (e.g., \citealt{Campana12}).  The X-ray
photon index in the 0.3--10~keV energy band is $\Gamma_X = 2.3 \pm
0.3$.

\begin{figure*}
\centering
\includegraphics[width=16cm]{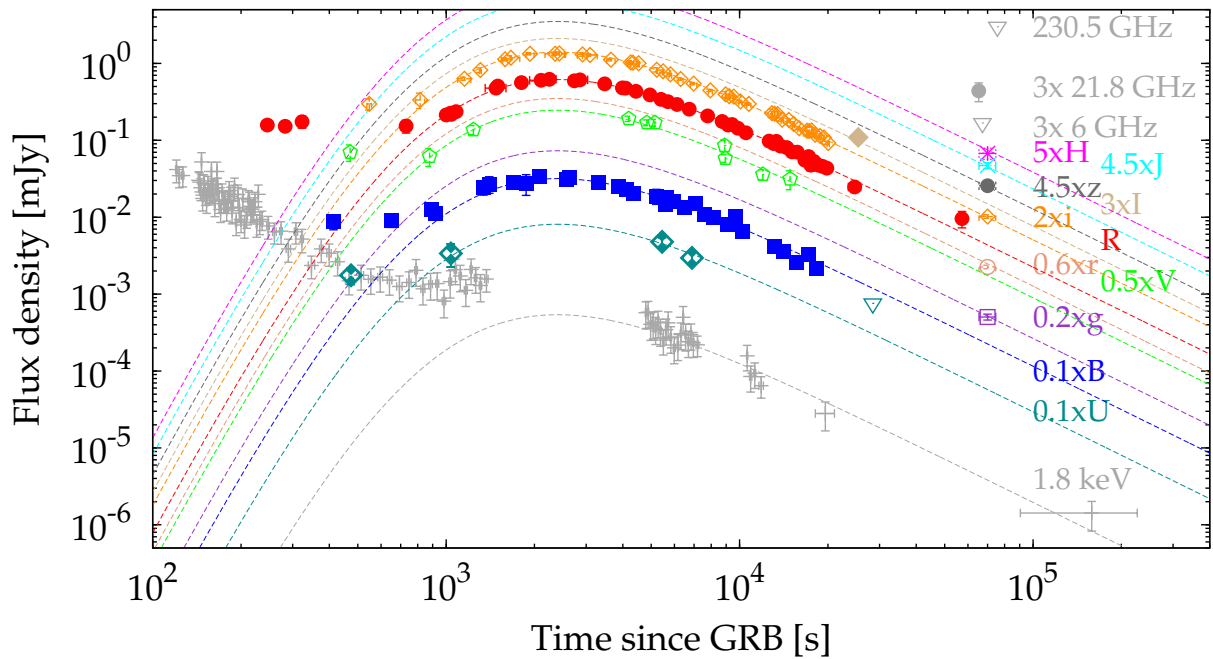}
\caption{Panchromatic light curve from radio to X-rays of the
early afterglow of GRB\,120404A. Upside down triangles show
upper limits. Normalisations have been rescaled for the sake of
clarity. Lines show the result of a simultaneous, achromatic fit
of the rebrightening. See text for further details.}
\label{f:lc_panchro}
\end{figure*}

The X-ray unabsorbed flux light curve was derived from the rate curve
by assuming the same counts-to-energy factor ($5.4 \times 10^{-11}$
erg~cm$^{-2}$~count$^{-1}$) obtained from the spectrum
described above.  This implicitly relies on the lack of strong
spectral evolution from $\sim 10$~ks onward; although such an
assumption cannot be proven due to the paucity of photons at late
times, this is in agreement with what is observed for most GRBs (e.g.,
\citealt{Evans09}).  Finally, the flux-density curve shown in
Figure~\ref{f:lc_panchro} was calculated at $1.8$~keV, the energy at
which the energy spectrum with $\beta_X = \Gamma_X - 1 = 1.3$ has the
same value as that averaged over the 0.3--10~keV range.
The X--ray light curve can be modelled ($\chi^2/{\rm dof}=61/100$) with a
double broken power--law with power--law indices $\alpha_1=2.28\pm0.24$,
$\alpha_2=-0.1\pm0.7$, $\alpha_3=1.8\pm0.3$, and break times $t_1=(540\pm120)$~s,
$t_2=(2480\pm460)$~s, respectively, in agreement with previous
reports \citep{Stratta12}.

\subsection{Optical and infrared data}
\label{sec:opt}

Both FTN and FTS carried out robotically triggered observations:
FTN observed from 4 to 75~min, while FTS observed from 17~min to
$5.6$~hr. The automatic identification of the afterglow by the
GRB pipeline LT--TRAP \citep{Guidorzi06} triggered the multi-filter
($BVRi'$) observation sequence. The optical afterglow position
is $\alpha$(J2000) $=15^{\rm h} 40^{\rm m} 02\fs30$,
$\delta$(J2000) $=+12^\circ 53\arcmin 06\farcs4$ with an error radius
of $0\farcs5$, consistent with the position determined by
{\it Swift}/UVOT \citep{Stratta12} and radio observations \citep{Zauderer12}.

%
%
Later observations were carried out with the $1.04$-m ARIES telescope
with the $RI$ filters. Observations started at 6.5~hr and last about
one hour through a sequence of four (three) individual frames in the
$R$ ($I$) filter 300-s exposure each. The afterglow is clearly
detected in the coadded frames for both filters.

We observed GRB\,120404A with the VLT/X-shooter spectrograph
at $15.9$~hr \citep{DElia12}. In particular, we obtained a photometric
estimate in the $R$ band from the 30-s exposure acquisition frame.

GROND observed GRB\,120404A from $18.2$ to $20.7$~hr simultaneously
with the $g'r'i'z'JHK$ filters. The afterglow was clearly detected in
all filters, except for $K$ for which an upper limit of $20.5$~mag was given.

Calibration of the $g'r'i'z'$ frames was performed against five field
stars of the Sloan Digital Sky Survey (SDSS; release 6).
Magnitudes in the Johnson-Cousins $BVRI$ filters for the same calibrating
stars were derived from the SDSS values using the transformations by
\citet{Jordi06}. For each filter the scatter in the zero point was
added in quadrature to the statistical uncertainty of each individual frame.
Both aperture and PSF photometry was systematically carried out using the
Starlink {\sc gaia} software\footnote{\texttt{http://starlink.jach.hawaii.edu/starlink}.},
making sure that both gave consistent results within the uncertainties.
In the case of VLT/X-shooter frames, the night was not photometric and
we could only use two faint field stars different from the five stars
mentioned above; for the acquisition frame, the zero-point was poorly
determined with an uncertainty of $0.3$~mag.
GROND $JHK$ filters were first calibrated against nearby 2MASS catalogue
stars and then converted to $AB$ magnitudes.

Magnitudes were finally converted into flux densities ($\mu$Jy) following
\citet{Fukugita95,Fukugita96}.
Table~\ref{tab:phot} reports the photometric set for all NIR/optical
data we collected. Magnitudes are corrected for airmass, while flux
densities are also corrected for Galactic reddening.

%
%

\subsubsection{Spectroscopy}
\label{sec:Gemini}
On April $4.57$ UT we observed the optical afterglow of GRB\,120404A with the Gemini--North
telescope and the GMOS camera \citep{Hook04}: we obtained $2\times900$~s spectra, using the
B600 grism with the $1\arcsec$\ slit (resolution of about $3.5$\,\AA)  centred at $6500$\,\AA, 
covering wavelengths $5000$--$8000$\,\AA.
The data were analysed using the standard {\tt GEMINI/GMOS} data analysis packages within the 
IRAF\footnote{IRAF is distributed by the National Optical 
Astronomy Observatory, which is operated by the Association for Research 
in Astronomy, Inc., under cooperative agreement with the National Science 
Foundation.} environment.
We performed flat--fielding, wavelength calibration (using a CuAr lamp spectra obtained
immediately after the science frames), and cosmic ray rejection using the {\tt lacos\_spec}
package \citep{Vandokkum01}.
A sky region close in the spatial direction, but unaffected by the spectral trace, was used for sky subtraction.
The two-dimensional spectra were then coadded. 
%
\begin{figure*}
\centering
\includegraphics[height=16cm,angle=90]{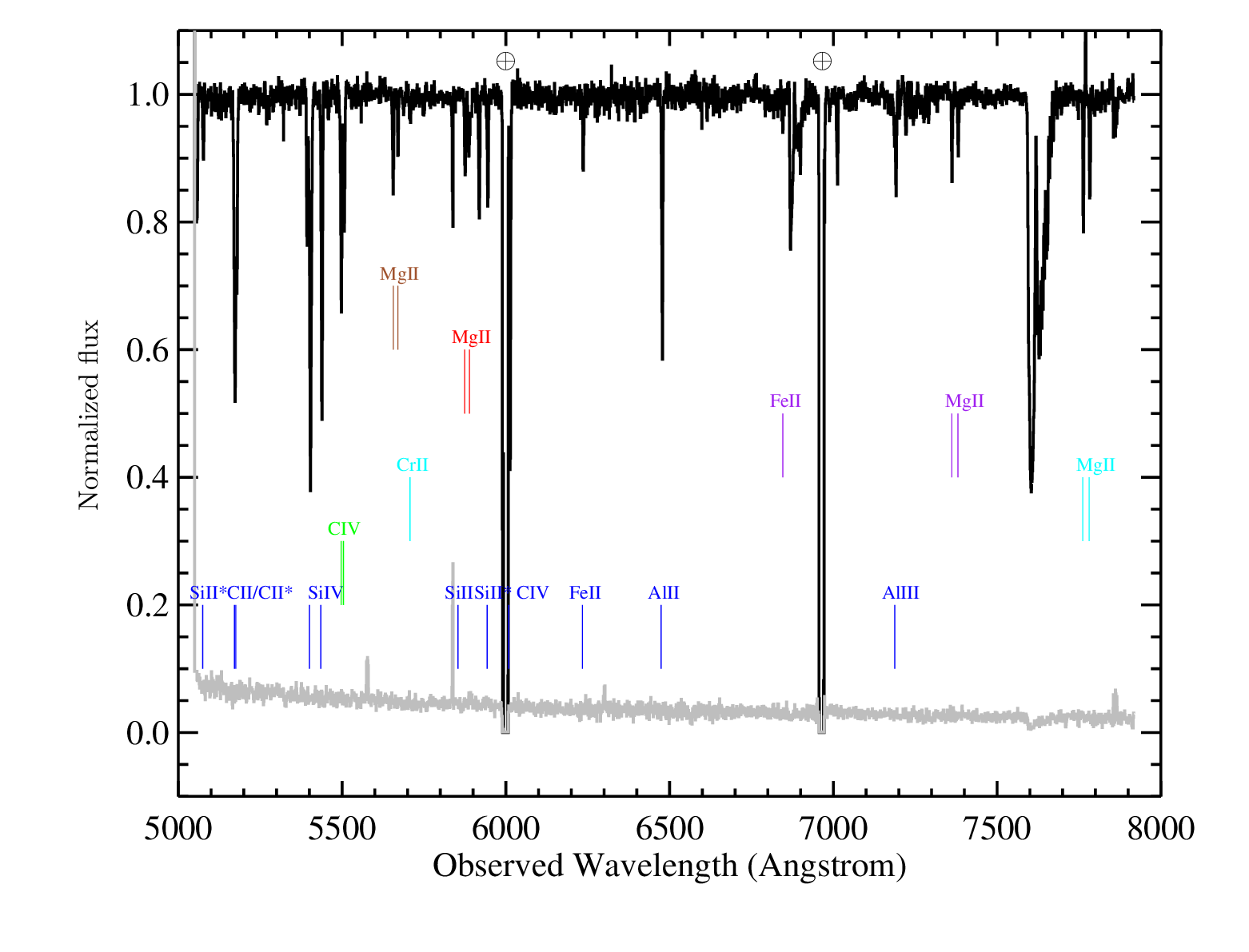}
\caption{Continuum normalised spectrum of GRB\,120404A observed with the Gemini--North (black is the spectrum
and grey is the error array associated with it): the main system at $z=2.8767$ (blue) is associated with the
GRB host galaxy and presents several low-- and high--ionised species as well as fine--structure transitions features
(e.g. SII*, CII, CII*, SiIV, CIV). In addition, four more systems have been identified, based on the
identification of metal lines: $z= 2.551$ (green, CIV), $z=1.776$ (cyan, CrII, MgII doublet),
$z=1.633$ (purple, FeII and MgII), $z=1.101$ (red, MgII), and $z=1.023$ (brown, MgII).}
\label{f:Gem}
\end{figure*}
%

Figure \ref{f:Gem} presents the final coadded spectrum, normalised to the continuum:
the afterglow spectrum presents a very complex series of absorption features. The main one,
indicated in blue, is associated to the GRB host galaxy (at $z=2.8767$) thanks to the identification
of low-- and high--ionised species as well as fine--structure transitions due to the UV radiation of
the GRB interacting with the interstellar medium \citep{Vreeswijk07,Prochaska06}.
In addition, we were able to identify an unusual set of intervening systems, based on MgII and CIV
doublets identifications, indicating a very complex line of sight (see also, GRB\,060418, \citealt{Vreeswijk07}).
These systems, at $z=2.551$, $1.776$, $1.633$, $1.101$, and $1.023$ are likely associated with galaxies
along the lines of sight.

\subsection{Radio and mm data}
\label{sec:radio}
We observed the position of GRB\,120404A with the Karl G. Jansky Very Large
Array \citep[JVLA][]{PerleyEVLA} at $21.8$~GHz (K band)  and $6.0$~GHz (C band)
beginning 2012 April 5 at 06:14:39~UT and with the Smithsonian Astrophysical
Observatory's Submillimeter Array (SMA; \citealt{HoSMA}) at $230.5$~GHz ($1.3$~mm)
beginning 2012 April 5 at 7:27:31~UT.  Observations are summarised in
Table~\ref{tab:radio}.  
A source of radio emission was detected at $21.8$~GHz with a flux of $87.6\pm 24.0~\mu$Jy
at a position of  $\alpha$ = 15:40:02.28 ($\pm$ 0.01) and $\delta$ = 12:53:06.1 ($\pm$ 0.1),
with 1$\sigma$ positional uncertainties.  This radio position is consistent with
both the {\it Swift}/XRT position \citep{Osborne12} and the UVOT position \citep{Stratta12}.  
No significant source of emission was detected at the position of GRB\,120404A at $6.0$~GHz
with the VLA to a 3$\sigma$ upper limit  of $33.6~\mu$Jy or at $230.5$~GHz with the SMA
to a 3$\sigma$ upper limit of $\sim$3~mJy.

VLA observations utilised the WIDAR correlator \citep{DoughertyPerleyEVLA}, with $1.024$~GHz
bandwidth in each of the upper and lower sidebands (eight intermediate frequencies per sideband,
each with 64, 2~MHz channels). At K band, we centred the frequency for each sideband at $19.1$~GHz
and $24.4$~GHz, with a mean frequency of $21.8$~GHz.  At C band, we centred the frequency for
each sideband at $4.9$~GHz and $7.0$~GHz, with a mean frequency of $6.0$~GHz.  In practice,
after flagging edge channels and excising RFI, we obtained a total continuum bandwidth of
$\sim$75$-$85\% (at C band where there is more RFI and K band, respectively).  
 
For VLA observations, we calibrated our bandpass and flux scaling using 3C286, and performed
gain calibrations with J1553+1256 ($3.3^{\circ}$ from GRB\,120404A). Observations were obtained
in the C configuration (maximum baseline $\sim3.4$~km), and the synthesised beam size is noted in Table \ref{tab:radio}. 
Reference pointing offsets were measured and applied prior to bandpass/flux observations
using 3C286 and prior to the gain calibrator/source observations using J1608+1029 at $8.4$~GHz,
according to standard VLA procedures.  Flagging, calibration, and imaging were 
performed using standard procedures in AIPS \citep{Greisen}.

SMA $230.5$~GHz observations were obtained in the very extended configuration, with baseline
lengths ranging from 103$-$476 m.  The synthesised beam size is noted in Table \ref{tab:radio}.
Neptune and Titan were utilised for flux measurements, 3C454.3 and 3C279 for bandpass, and
1550+054 and 1540+147 for gain calibration. Data reduction was performed using standard MIRIDL
and MIRIAD procedures.

\section{Modelling and interpretation}
\label{sec:mod}

\subsection{Broadband fitting}
\label{sec:broad}
We modelled the multi--filter light curves simultaneously by imposing common
power--law indices, given the apparent lack of evidence for strong chromatic
evolution during the rebrightening. We adopted the same approach as for past
events (e.g., see \citealt{Guidorzi11b}) with important changes: in the previous
treatment, the different normalisations of each filters, initially treated
as independent parameters, were then used to construct a SED.
In this case we adopted a more general approach, since we fitted both the temporal
and spectral dependence of the flux density at different wavelengths
simultaneously.
We assumed the SED to be described by a simple power--law model,
$F(\nu,t)\propto\nu^{-\beta_0}\times10^{-0.4\,A(\nu)}$, where the term $A(\nu)$
accounts for the rest--frame dust extinction as modelled according to three
different (SMC, LMC, MW) profiles in the \citet{Pei92} parametrisation.
The temporal behaviour was modelled in the time interval from 800 to $2\times10^5$~s,
i.e. from the rebrightening onset.
The complete model describing the temporal evolution of flux densities at all optical
wavelengths is given in eq.~(\ref{eq1}),
\begin{eqnarray}
\displaystyle F(\nu,t) & = & F_{15}\ \nu_{15}^{-\beta_0}\,10^{-0.4\,A(\nu)}
\times\nonumber\\
& & \times \Bigg[ \frac{1 - \alpha_1/\alpha_2}{\Big(\frac{t}{t_{\rm p}}\Big)^{n\alpha_1} +
\Big(\frac{t}{t_{\rm p}}\Big)^{n\alpha_2} \Big(-\frac{\alpha_1}{\alpha_2}\Big)} \Bigg]^{1/n}\;,
\label{eq1}
\end{eqnarray}
where the free parameters are $F_{15}$, i.e. the dust--unextinguished flux density at
$\nu=10^{15}$~Hz ($\nu_{15}=\nu/10^{15}$~Hz) at the peak time ($t=t_{\rm p}$), the spectral index
$\beta_0$, the extinction $A_V$ incorporated in the term $A(\nu)$, the rise $(\alpha_1<0)$
and decay ($\alpha_2>0$) power--law indices, the peak time $t_p$, the smoothness parameter $n$.
Frequencies are expressed in the GRB rest--frame. We chose to use the peak time rather
than the break time (e.g., see the parameter $t_{\rm b1}$ in eq.~1 of \citealt{Guidorzi11b}),
as the free parameter, since $t_{\rm p}$ is the interesting parameter and its uncertainty does not
have to be calculated taking into account the complicated covariance with other free parameters
as it is required for derived (i.e., not free) quantities.
Best--fitting parameters were found by minimising the total $\chi^2$, as expressed in
eq.~(\ref{eq2}),
\begin{equation}
\chi^2(F_{15},\beta_0,A_V,\alpha_1,\alpha_2,t_{\rm p},n)\ =\ \sum_{k,i}\
\Big(\frac{F(\nu_k,t_{k,i})-F_{\nu_k,i}}{\sigma_{\nu_k,i}}\Big)^2\;,
\label{eq2}
\end{equation}
where $F_{\nu_k,i}$ and $\sigma_{\nu_k,i}$ are the measured flux density and uncertainty
for $\nu=\nu_k$ at the time $t=t_{k,i}$.
The observed filter $U$ has an effective rest--frame wavelength of $890$~\AA, i.e. bluewards
of the Lyman limit of $912$~\AA. Its flux density is thus heavily suppressed by the neutral
hydrogen along the sightline. Because of this, we let the normalisation
constant for the $U$ filter to be independently determined by the fitting procedure.

The first three lines in Table~\ref{tab:LCSED} report the results obtained by fitting the
optical data alone with the three different dust extinction profiles.
Although all of the extinction profiles yield formally acceptable $\chi^2$ values, in the
following we show that only the MW profile provides a plausible and self--consistent
description of the SED. All models provide identical temporal evolution of the light curves,
the only discrepancies concerning the spectral parameters.
As is often the case, the rise slope is determined with large uncertainty, whereas the
decay slope is more accurately measured, $\alpha_2=1.9\pm0.1$. The peak time essentially
remains the same, around $2.4\pm0.6$~ks, regardless of the model adopted.
\begin{table*}
\centering
\caption{Spectral and temporal best--fitting parameters describing the evolution
of the flux densities at optical and X--ray wavelengths.}
\label{tab:LCSED}
\begin{tabular}{llcccccccc}
\hline
Data & Dust & $F_{15}^{\mathrm{(a)}}$   & $\beta_o$ & $A_V^{\mathrm{(b)}}$  & $\alpha_1$  & $t_{\rm p}$   & $\alpha_2$ & $n$ & $\chi^2/$dof\\
     &      & (mJy)     &            & (mag) &             &    (s)        &           &     &             \\
\hline
optical & MW & $1.89\pm0.09$ & $1.05\pm0.09$ & $0.22\pm0.05$ & $-7.5_{-4.6}^{+5.2}$ & $2395\pm55$ & $1.89\pm0.06$ & $0.16_{-0.08}^{+0.16}$ & $89/124$\\
optical & LMC & $1.93\pm0.12$ & $0.39_{-0.22}^{+0.25}$ & $0.32\pm0.08$ & $-6.6\pm3.7$ & $2400\pm55$ & $1.86\pm0.06$ & $0.19_{-0.09}^{+0.15}$ & $98/124$\\
optical & SMC & $1.67\pm0.20$ & $-0.23_{-0.37}^{+0.68}$ & $0.32\pm0.09$ & $-6.4\pm3.7$ & $2401\pm55$ & $1.86\pm0.06$ & $0.19_{-0.09}^{+0.16}$ & $125/124$\\
opt--X & MW & $1.89\pm0.06$ & $1.05\pm0.03$ & $0.22\pm0.04$ & $-7.2_{-4.3}^{+4.7}$ & $2397\pm55$ & $1.88\pm0.06$ & $0.16_{-0.08}^{+0.16}$ & $103/150$\\
opt--X & LMC & $1.57\pm0.05$ & $1.04\pm0.02$ & $0.12\pm0.03$ & $-7.2_{-4.2}^{+4.7}$ & $2399\pm55$ & $1.89\pm0.05$ & $0.17_{-0.09}^{+0.16}$ & $172/150$\\
opt--X & SMC & $1.42\pm0.05$ & $1.04\pm0.02$ & $0.07\pm0.02$ & $-7.5\pm4.5$ & $2397\pm55$ & $1.90\pm0.05$ & $0.16_{-0.08}^{+0.16}$ & $167/150$\\
\hline
\end{tabular}
\begin{list}{}{}
\item[$^{\mathrm{(a)}}$]{Flux density at the rest--frame frequency of $10^{15}$~Hz, at peak
and corrected for dust extinction.}
\item[$^{\mathrm{(b)}}$]{Rest--frame quantity.}
\end{list}
\end{table*}

While in Section~\ref{sec:optX_SED} we modelled a detailed optical--X--ray SED taking into account
the X-ray spectral shape itself, here we preliminarily added the X--ray flux history obtained
by assuming a constant count--to--flux conversion. We determined the reference energy
$\hat{E}=1.8$~keV, i.e. the energy at which the flux density is the same as the average one
in the XRT passband $0.3$--$10$~keV for a power--law spectrum with $\Gamma_X=2.3$ obtained
in Section~\ref{sec:xray}. Excluding the initial steep decay ($t<800$~s), which clearly has
a different origin from the subsequent emission, X--rays exhibit the same temporal behaviour
as the optical photons. This justifies a common fit. For the same reason, we also
exclude the presence of any break frequency between optical and X--rays, so a simple power--law
appears to be the only plausible spectrum. Fitting all data sets together, one obtains
almost identical results for the MW extinction profile, which is still the best model by far,
as reported in the last three lines of Table~\ref{tab:LCSED}.
For the two remaining profiles, forcing no break between optical and X--rays clearly
changes the spectral index from the corresponding previous cases where X-ray data had not been considered.
Although these models cannot be rejected solely because of their $\chi^2$ values,
the modelling obtained assuming a MW profile offers by far the best, and most self--consistent
description of our data, thus lending support to the evidence for the presence of a 2175~\AA\ 
bump. The resulting MW--profile based model for each light curve is shown together with data
in Figure~\ref{f:lc_panchro}.

From the accurate spectral and temporal modelling we can estimate the total energy released
in the optical--to--X-ray frequency range during the rebrightening from $800$~s on, $E_{\rm reb}$,
properly corrected for dust extinction.
Strictly speaking, since the low--energy part of the SED as well as the flux at
early times ($t<800$~s) are poorly known, our estimate should be taken as a lower limit.
However, taking into account the uncertainty on $\alpha_1$ and extrapolating the power--law spectrum
to much lower frequencies, the result does not change by more than a factor of two.
\begin{eqnarray}
E_{\rm reb} & \ga & \frac{4\pi\,D_{\rm L}^2}{1+z}\ \int_{\nu_{H}}^{\nu_{\rm x}}d\nu\ \int_{800\,{\rm s}}^{+\infty}dt\ F(\nu,t)10^{0.4A(\nu)} \\\nonumber
& = & 15\,F_{15}\Big(\frac{\nu_{\rm x,15}^{1-\beta}-\nu_{\rm H,15}^{1-\beta}}{1-\beta}\Big)\ 
\textrm{erg}\ =\  2\times10^{52}\ \textrm{erg}\;,
\label{eq:energy_reb}
\end{eqnarray}
where $D_{\rm L}=7.6\times10^{28}$~cm is the luminosity distance.
Hence, the energy released during the rebrightening is a non--negligible fraction of the isotropic--equivalent
one released in the prompt emission, $E_{\gamma,{\rm iso}}=(9\pm4)\times10^{52}$~erg (Section~\ref{sec:gamma}).

\subsubsection{Evidence for chromaticity}
\label{sec:achrom}
Although a simple, achromatic model for the rebrightening and subsequent decay was shown to
provide an acceptable description, we investigated whether there exists evidence for chromatic
evolution, by allowing different peak times for the light curves at different wavelengths.
To this aim, we applied the same fitting procedure as in Section~\ref{sec:broad}, but
allowing independent peak times for the best sampled filters: $i'$, $R$, $V$, $B$, and X--ray.
For the remaining filters we used the sample peak time as that of the closest-in-frequency
filter among those treated as free parameters.
Limiting to the best--fitting case given by the MW dust extinction one, the total $\chi^2$/dof
improved from that obtained in the strictly achromatic case, 103/150, to 89/146.
Such values for the total $\chi^2$, being smaller than 1, probably reflect that uncertainties
on individual measures have likely been overestimated following a conservative approach.
Formally, the p--value according to the additive F--test is $4\times10^{-4}$.
However, the small  $\chi^2$ values suggest a more conservative F--test calculation
assuming a unitary reduced $\chi^2$ for the chromatic model, which yields a p--value of $1.0$~\%.
We therefore conclude that there is evidence for chromatic evolution with $\la 1$\% confidence.
What is more, the peak time as a function of wavelength follows a precise
trend: the higher the effective frequency, the earlier the light curve
seems to peak, as reported in Table~\ref{tab:tp_vs_freq}.
Should the improvement be entirely due to chance, one would expect no such trend between
peak time and frequency.
\begin{table}
\centering
  \caption{Peak time as a function of rest--frame frequency.}
  \label{tab:tp_vs_freq}
  \begin{tabular}{ccl}
\hline
Observed filter & Rest--frame $\nu_{\rm eff}$ & Peak time $t_{\rm p}$\\
                & ($10^{15}$~Hz)              & (s)\\\hline
$i'$ & $1.56$ & $2420\pm40$\\
$R$  & $1.81$ & $2360\pm60$\\
$V$  & $2.14$ & $2350\pm140$\\
$B$  & $2.67$ & $2220\pm100$\\ 
X    & $1686$ & $1310_{-150}^{+170}$\\
\hline
\end{tabular}
\end{table}
Modelling this dependence with a power--law, $\nu_{\rm eff}\propto t^{-\delta}$,
yields $\delta=12\pm4$, where the time origin was fixed to the GRB trigger time.
Interpreting this as the crossing of a given break frequency through different
filters at different times, as one would expect for the synchrotron spectrum evolution,
the temporal dependence is too strong to match any theoretical expectation,
unless one resets the time origin. Ignoring the X--ray band, the same index
is poorly constrained, $\delta=6\pm3$, which still implies a strong evolution.
Overall, the evidence for a time lag in the peak as a function of frequency
cannot be considered compelling, but surely plausible and likely.

\subsubsection{Optical-to-X-ray spectral energy distribution}
\label{sec:optX_SED}
%
\begin{figure*}
\centering
\includegraphics[width=8.2cm]{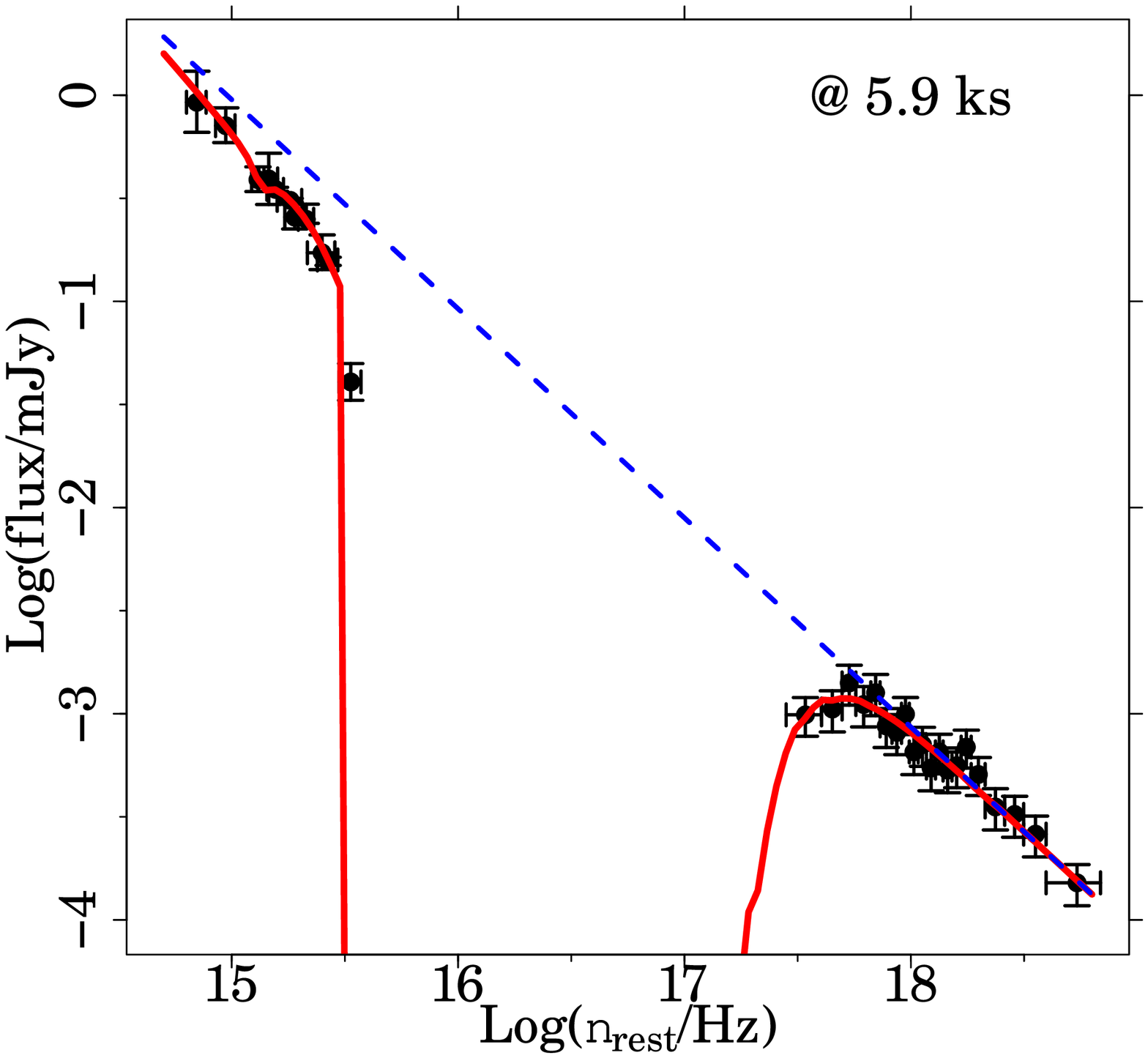}
\includegraphics[width=8.2cm]{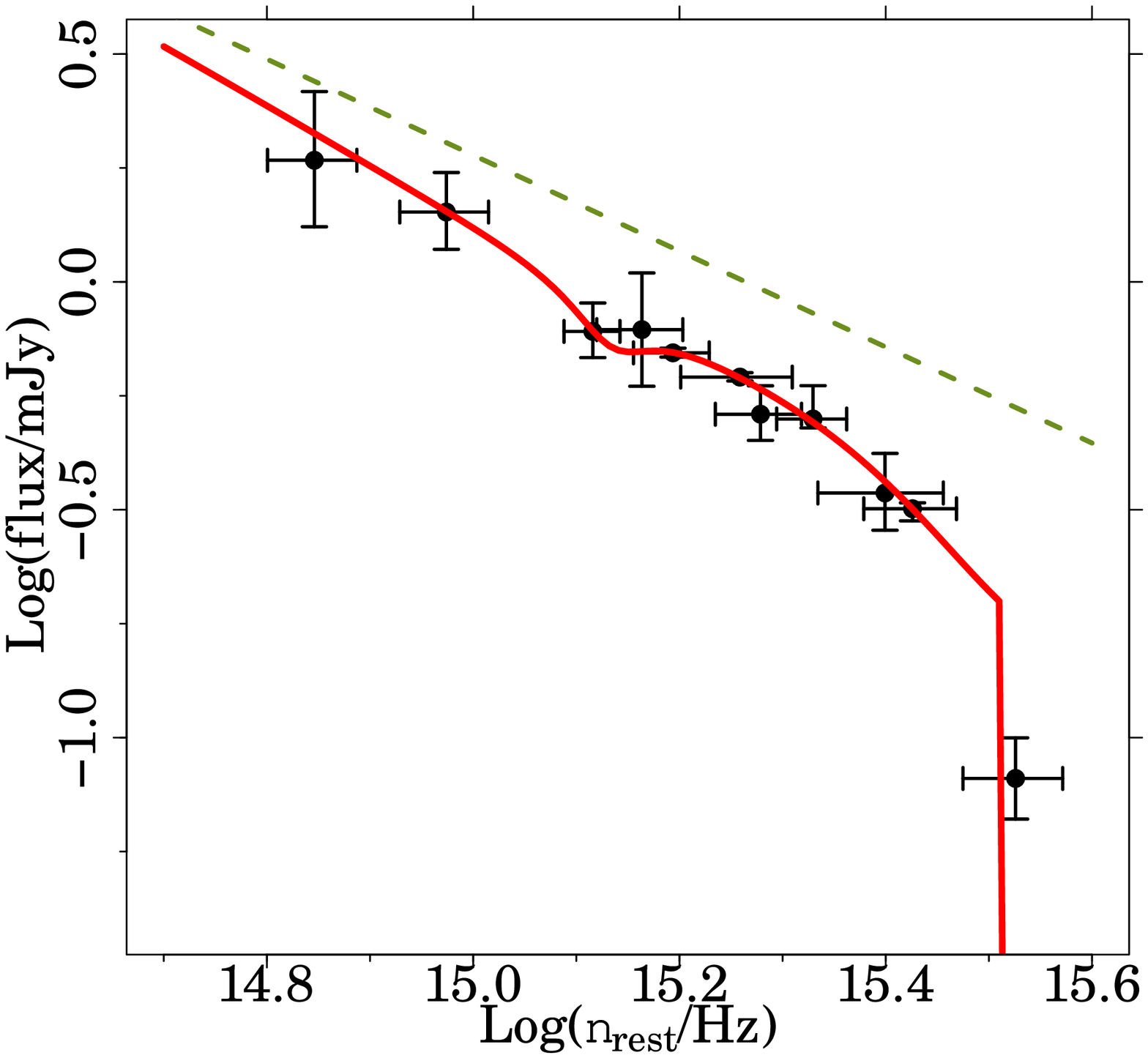}
\caption{{\em Left panel}. Rest--frame optical--X SED at $\hat{t}=5.9$~ks. The solid line
shows the best-fitting model obtained with a MW extinction profile and a simple
power--law with index $\beta=1.01\pm0.03$. The dashed line shows the SED one
would have observed in the absence of dust. {\em Right panel}. Close--in of the
optical points.}
\label{f:seds}
\end{figure*}
%
Although XRT could not collect data around the optical peak, still the available data support
the view that, after the initial steep decay, the X--ray flux underwent the same temporal
rebrightening followed by an analogous power--law decay.
We therefore accumulated an X--ray energy spectrum right after the optical peak, from
$4.7$ to $7.3$~ks, i.e. when the final power--law decay with $\alpha_2=1.9$ already set in.
The reference time when the instantaneous flux is the same as the average one over
the above time interval, is found to be $\hat{t}=5.9$~ks.
To determine the optical flux densities at each wavelength, we ran the multi--filter
procedure of Section~\ref{sec:broad}. However, we did not model the spectral parameters,
but we introduced an independent normalisation term each light curve to be freely
determined, in the same fashion as we used to do for previous GRBs (e.g., \citealt{Guidorzi11b}).
Not surprisingly, the temporal parameters describing the light curve evolution did not
change (Table~\ref{tab:LCSED}). The best--fitting normalisations at each filter, expressed
as flux densities at peak, are reported in Table~\ref{tab:normSED}.
\begin{table}
\centering
  \caption{Best--fitting flux densities at peak.}
  \label{tab:normSED}
  \begin{tabular}{lcc}
\hline
Parameter & Value & Unit\\
\hline
$F_H$          &  $1850_{-530}^{+750}$ & $\mu$Jy\\
$F_J$          &  $1420_{-250}^{+300}$ & $\mu$Jy\\
$F_z$          &  $778_{-100}^{+114}$ & $\mu$Jy\\
$F_I$          &  $785_{-200}^{+260}$ & $\mu$Jy\\
$F_i$          &  $699\pm16$ & $\mu$Jy\\
$F_R$          &  $618\pm13$ & $\mu$Jy\\
$F_r$          &  $512\pm8$ & $\mu$Jy\\
$F_V$          &  $500_{-40}^{+43}$ & $\mu$Jy\\
$F_g$          &  $344_{-61}^{+73}$ & $\mu$Jy\\
$F_B$          &  $318\pm14$ & $\mu$Jy\\
$F_U$          &  $81_{-15}^{+19}$ & $\mu$Jy\\
$\chi^2/{\rm dof}$   &  $83/117$\\
\hline
\end{tabular}
\end{table}
Optical flux densities at the X--ray spectrum reference time $\hat{t}$ were
calculated simply by rescaling the corresponding peak values using the
temporal model in Table~\ref{tab:LCSED}, which yielded a factor of $0.50$.
We thus constructed an optical--X SED at $\hat{t}$ by rescaling the
optical flux densities. Given the same temporal decay exhibited by optical
and X--ray profiles, no break frequency in between is to be expected,
consistently with the simple power--law model adopted in Section~\ref{sec:broad}.

In addition to modelling the dust extinction, we also had to account for
the photoelectric absorption which suppressed the soft X--ray flux.
We modelled this using the photoelectric cross section as parametrised by
\citet{Morrison83}. The amount of gas responsible for this absorption was
modelled in terms of neutral hydrogen column density evaluated in the GRB
rest--frame, $N_{\rm H}$, assuming solar abundances. The Galactic contribution
was accounted for separately.

The only acceptable model is that obtained assuming a MW extinction profile.
Its best--fitting parameters were: $\beta=1.01\pm0.03$, $A_V=0.24\pm0.07$~mag,
$N_{\rm H}=5.2_{-1.7}^{+2.6}\times10^{21}$~cm$^{-2}$, $\chi^2/{\rm dof}=32.4/27$, with
a null hypothesis probability of 22\% (Table~\ref{tab:LCSED}).
This result is fully compatible with what is obtained adopting the same extinction
profile when we fitted the optical data alone (Section~\ref{sec:broad}).
The result is shown in Figure~\ref{f:seds}.

\subsection{The standard afterglow model}
\label{sec:broadband}
In the context of the standard afterglow model
(see, e.g., \citealt{Meszaros06,Vaneerten13b,Gao13} for reviews), a population of
shock--accelerated electrons cools through synchrotron emission,
 resulting in spectra and light curves that are characterised by power--law segments,
which join at given break frequencies.
The electron energy distribution is assumed to be $dN/d\gamma \propto \gamma^{-p}$ ($\gamma > \gamma_m$).
Typical values for $p$ found from GRB afterglow modelling range
between $2$ and $3$, in broad agreement with theoretical
expectations (e.g., \citealt{Spitkovsky08}).
At sufficiently late times the afterglow emission is dominated by the forward
shock (FS), i.e. the emission due to the shocked interstellar medium, because the emission of
the reverse shock - which propagates within the ejecta - is short--lived.
At such times, the corresponding observed spectral and temporal decay indices for GRB\,120404A
are $\beta=1.0$ and $\alpha_2=1.9$ (Table~\ref{tab:LCSED}), with no apparent break
from optical through X--rays (Fig.~\ref{f:seds}).
Using $\nu_{\rm m,f}$ and $\nu_{\rm c,f}$ for synchrotron injection and cooling
frequencies respectively associated with the FS, the most plausible scenario is the slow
cooling regime at $\nu_{\rm m,f}<\nu_{\rm opt}<\nu_{\rm x}<\nu_{\rm c,f}$, for which $\beta=(p-1)/2$
yields $p=3$. The decay index depends on the density profile of the circumstellar medium density
such that  for a homogeneous (wind) medium  $\alpha=3(p-1)/4=1.5$ ($\alpha=(3p-1)/4=2$). 

Thus, simple analytical expectations show that a density profile more akin to a wind
could be compatible with the observed spectral and temporal afterglow properties at late times.
The closer in time to the initial prompt emission, the more complicated is
the overall description of the observed radiation, due to multiple overlapping
components from co-located or distinct emitting regions: e.g., an emitting reverse shock,  energy
injection due to on--going activity of the inner engine, or the onset of the afterglow itself
due to the deceleration of the ejecta by the surrounding medium.

Large and accurate broadband data sets for a given GRB afterglow hold the potential
to self--consistently constrain the geometry and dynamics of the relativistic outflow,
the density profile of the circumstellar medium, as well as the detailed microphysics of
the shock acceleration of electrons and local magnetic field generation.
Here we show that even at late time, when most of the jet energy has already been
transferred to the shocked ambient medium, a realistic and detailed physical description
requires comparably realistic modelling.
To do this, we adopt the model developed by \citet{Vaneerten12b}. In addition, we model,
separately, the early time emission ($t<800$~s) when the afterglow is likely to be
dominated by reverse-shock emission.

%
\begin{figure*}
\centering
\includegraphics[width=16cm]{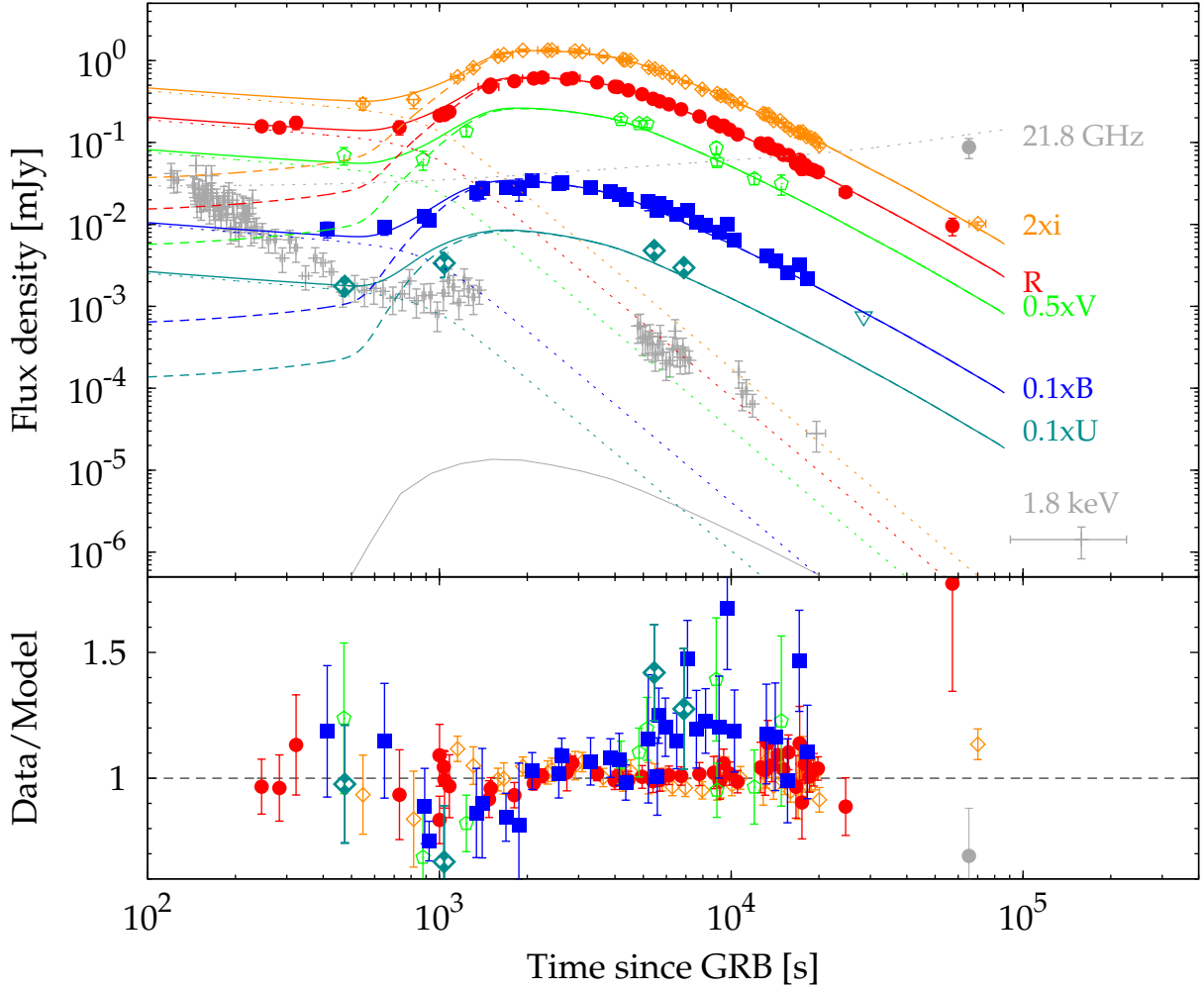}
\caption{{\em Top panel}: light curves from radio to X-rays of the early afterglow of
GRB\,120404A. Here only the best sampled filters are shown together
with the radio detection at $21.8$~GHz.
The models superposed to each data set are the synthetic light curves
obtained with the {\sc boxfit} code which best fit the corresponding
data set. The continuously refreshed RS contribution (dotted line) is visible at
early times, while the FS (dashed line) takes over at $t\ga 10^3$~s.
{\em Bottom panel}: fractional residuals.}
\label{f:boxfit}
\end{figure*}
%
\subsection{Relativistic shock physics and jet geometry}
\label{sec:boxfit}
The interpretation of the late time broadband rebrightening in terms of radiation
coming from an off--axis jet which finally reaches the observer proved successful
in a number of similar cases, such as GRB\,081028 \citep{Margutti10}.
We investigated the plausibility of this scenario for GRB\,120404A by fitting our
multi-frequency data set with the {\sc boxfit}
code.\footnote{\texttt{http://cosmo.nyu.edu/afterglowlibrary/index.html}}
This code assumes a homogeneous jet with sharp edges ploughing into a constant
density medium; it is possible to calculate afterglow light curves and spectra due
to synchrotron radiation at any observer time and
frequency and the code performs data fitting with the downhill simplex method combined
with simulated annealing. The blast wave dynamics have been calculated for
19 high--resolution, 2--D jet simulations performed with the relativistic adaptive
mesh (RAM) parallel relativistic hydrodynamical (RHD) code. Exploiting the scale
invariance of different jets with different energies and circumburst densities,
the code calculates spectra and light curves by solving the linear radiative transfer
equations including synchrotron self--absorption.
At the cost of a relatively limited amount of computational time, it properly accounts
for features such as jet decollimation, inhomogeneity along the shock front, and its
late transition to non--relativistic regime.
The free parameters include the jet geometry, the energetics and circumburst properties,
the released energy, and the microphysics parameters which determine the basic properties of
the synchrotron radiation caused by the relativistic shocks.

We fitted our broadband data set with the ``Fermi'' hybrid server for high performance
computing of the University of Ferrara, equipped with 188~GB DDR3 of
RAM memory.\footnote{\texttt{http://fermi.unife.it}}
The free parameters were the jet half--opening angle $\theta_0$, the isotropic--equivalent
total released energy $E_{\rm iso}$, the circumburst number density $n$, the viewing angle
$\theta_{\rm obs}$, the electron energy distribution index $p$, the fractions of internal
energy going into magnetic fields and accelerated electrons, $\epsilon_B$ and $\epsilon_e$,
respectively, and the fraction of accelerated electrons, $\xi_N$.
The code assumes that the fireball energy has already been transferred to the ISM,
since it makes use of the Blandford--McKee solution (BM; \citealt{BM76}) as long as the
Lorentz factor of the shocked interstellar matter is high enough. Consequently, the free
parameters exclusively concern the propagation and the radiation of the FS.
\footnote{We made sure that results did not depend appreciably on the adopted value for the
{\tt BM\_start} parameter, by choosing a range of plausible values for it, as recommended
by the code's authors.}

First, we corrected all the optical flux densities for a factor corresponding
to how much the flux in each filter had been suppressed due to the local dust, because the code
does not account for it. In this case, thanks to the robust estimate we obtained for
the dust content (Section~\ref{sec:optX_SED}), this should not introduce a big source of
uncertainty and, in any case, the correction was within a factor of 2 for all cases,
except for the $U$ filter.
We obtained a remarkably good result for all the radio and optical data points
starting from the onset of the rebrightening onwards, i.e. at $t>800$~s. However,
in none of the allowed cases the best-fitting result could provide a successful
match of the X--ray data, for which the best--fitting solution underestimates the
X--ray flux by more than a factor of 10.
The best--fitting parameters obtained in this case are the following
(Table~\ref{tab:boxfit}):  $E_{\rm iso,53}=1.8$, $n_0=86$,
$\theta_0=26^{\circ}$, $\theta_{\rm obs}=25^{\circ}$, $p=3.6$,
$\epsilon_{B,-4}=4.7$, $\epsilon_{e,-2}=8.3$, with $\xi_N$ fixed to $1.0$
($\chi^2/{\rm dof}=451/162$), where $\epsilon_{B,-4}=\epsilon_B/10^{-4}$, 
$\epsilon_{e,-2}=\epsilon_e/10^{-2}$.

\subsubsection{X-ray excess and local cooling}
A possible explanation for the underestimated X--ray flux likely lies in the global
cooling time approximation, which is known to systematically underestimate the flux
beyond the cooling break \citep{Vaneerten10a}. The model adopted by the fitting
code assumes a common synchrotron cooling time for all the fluid cells which are
contributing to the observed spectrum. In reality, electrons are shock--accelerated
at the blast wave front at different times for different fluid cells.
As a result, the cooling time should be calculated from the time at which each local
fluid element is shocked, which has a local dependence. The cooling frequency should
be calculated with a much higher spatial resolution than the fluid scale.
In the global cooling approximation, the plasma is treated as a whole rather than locally.
Consequently, the flux above the cooling frequency is systematically underestimated by
a factor, which can be of the order of 10 or more, as shown by \citet{Vaneerten10a}.
Indeed, the X--ray flux curve matches the overall observed behaviour and the shift
required for a good match can in principle be entirely explained replacing the global
cooling with the more realistic local cooling approximation.

Under the assumption that the mismatch between model and X--ray data is accounted
for by the global cooling approximation, we ignored X--rays and fitted the remaining
data set. The result is shown in Fig.~\ref{f:boxfit} (dashed lines) and corresponds
to the following set of best--fitting parameters: $E_{\rm iso,53}=1.9_{-0.1}^{+0.7}$,
$n_0=240_{-90}^{+10}$, $\theta_0=23.1_{-4.1}^{+0.8}$~degrees, $\theta_{\rm obs}=(0.93\pm0.01)\,\theta_0$,
$p=3.8\pm0.1$, $\epsilon_{B,-4}=2.4_{-0.3}^{+0.1}$, $\epsilon_{e,-2}=9.3_{-3.4}^{+0.5}$,
$\xi_N=1.0_{-0.4}$ ($\chi^2/{\rm dof}=173/122$). The parameters' uncertainties were
calculated through the Monte Carlo procedure implemented in the code after determining
the partial derivatives around the minimum.
Although formally the quality of the fit is still poor (null hypothesis probability
of $0.2$\%), the overall capability of the code to reproduce the multi--band light curves
from radio to UV is noteworthy. In particular, the fit residuals amount to a few \%
for the most accurate data points, whose uncertainties are comparably small.

\subsection{The nature of the early optical emission}
\label{sec:early}
The early optical emission is likely to be produced by a separate emission process
from that of the later-time emission given the sharp change in the temporal evolution
after $\sim$800~s. The flux is roughly consistent with being constant with time.
We first tried to characterise this emission phenomenologically.
The data points covering this part are too few for fitting
eq.~(\ref{eq1}) with the same free parameters as used in Section~\ref{sec:broad}.
We therefore fixed the dust content to the previously determined value of $A_V=0.22$~mag.
Nevertheless, the spectral index was poorly determined as $\beta_0=0.6\pm0.6$, i.e. roughly
consistent with the later value of $\sim1.0$. The flux density at the rest--frame frequency of
$10^{15}$~Hz is $250_{-90}^{+150}$~$\mu$Jy ($\chi^2/{\rm dof}=1.0/5$).

We therefore instead tried a more physically motivated approach.
One of the most natural and least ad--hoc possibilities
is the presence of a short--lived RS accompanying the FS.
The RS is to be expected whenever given conditions are fulfilled.
This is the case when the magnetisation degree
of the fireball $\sigma$, defined as the ratio of electromagnetic and
kinetic energy density of the ejecta, is neither $\sigma\ll 1$ nor $\sigma>1$
\citep{ZK05}. The various possible combinations of RS+FS light curves depend
on the value of the synchrotron frequencies at deceleration of both shocks,
$\nu_{\rm m,r}(t_{\rm d})$ and $\nu_{\rm m,f}(t_{\rm d})$ respectively,
with respect to the observed frequency \citep{Zhang03,Gomboc09,Harrison13}.
In particular, in some cases the presence of a single peak or, more generally,
the lack of evidence for a RS contribution, is explained in the context of
the low--frequency model by relatively small microphysics parameters $\epsilon_e$
and $\epsilon_B$ which determine correspondingly small values for $\nu_{\rm m,r}$
and $\nu_{\rm m,f}$ \citep{Mundell07,Melandri10,Guidorzi11b}.

The value of the dimensionless parameter $\xi_0=(l/\Delta_0)^{1/2}\,\Gamma_0^{-4/3}$ determines
the evolution of the RS propagating through the ejecta, where $l=(3E_{\rm iso}/4\pi\,m_p\,n\,c^2)^{1/3}$
is the Sedov length, $\Delta_0\simeq c\,T_{90}/(1+z)$ is the shell radial width in the coasting phase
when it moves with a Lorentz factor $\Gamma_0$ before the deceleration sets in.
From the FS modelling obtained in Section~\ref{sec:boxfit} we calculate $l=5\times10^{17}$~cm.
The deceleration time $t_{\rm d}$ must occur before $200$~s, so it is possible to derive directly
a lower limit to $\Gamma_0$,
\begin{equation}
t_{\rm d} = T_{90} + \frac{0.2}{\Gamma_0^{8/3}}\,\frac{l}{c}\,(1+z) < 200\ \textrm{s}\;,
\label{eq:td}
\end{equation}
where we used the numerical result $t_{\rm d}=(0.2 + \xi_0^{-2})\,l/c\,\Gamma_0^{8/3}$ \citep{Harrison13}.
The condition on the initial bulk Lorentz factor is $\Gamma_0>71$. This constrains the
shell regime to be $\xi_0<4$, which corresponds to the intermediate/thick shell regime
\citep{Kobayashi99,Harrison13}.

Using this framework, we examine two possible interpretations of the early ($\la 800$~s) optical emission.

\subsubsection{A short--lived reverse shock}
\label{sec:boxfit_RS}
If the early time emission originates from the RS emission alone, one may explain
the shallow--to--steep evolution as the passage of the RS typical frequency $\nu_{\rm m,r}$
\citep{Kobayashi00a}.
In the $i'$ band we also see the passage of the FS component, so using these two times
it is possible to infer estimates of $\Gamma_0$, of the parameter $\xi_0$, and of
magnetisation $R_{\rm B}=\epsilon_{B,{\rm r}}/\epsilon_{B,{\rm f}}$.

In the thick shell case, slow cooling regime, and frequency range
$\nu<\nu_{\rm m,r}(t_{\rm d})$, after the deceleration the flux is expected
to decrease with a slope of $\alpha_{\rm r,1}=17/36$ until
$\nu_{\rm m,r}\propto t^{-73/48}$ crosses the observed band, after which the slope
steepens to $(73p+21)/96$ \citep{Kobayashi00a}.
We fitted the early dust--corrected $UBVRi'$ fluxes imposing the afore--mentioned expected temporal
evolution for the flux, that for $\nu_{\rm m,r}$, the spectral slope of $1/3$ at $\nu<\nu_{\rm m,r}$
\citep{Sari98}, and left free to vary only two parameters, the crossing time of $\nu_{\rm m,r}$
through a given band (we chose the best sampled $i'$), $t_{\rm r,m,i}$, and the corresponding
flux density, $F_{\rm r,m,i}$. We fitted the observed flux densities removed of
the contribution of the FS as modelled with the {\sc boxfit} code.
In spite of the very few additional degrees of freedom, the result is satisfactory.
The overall quality of the RS+FS model of the entire data
set, excluding the X--ray band, improves to $\chi^2/{\rm dof}=178/129$ (p--value of 
0.3\%; Table~\ref{tab:boxfit}), so basically equivalent to the refreshed shock modelling
discussed in Section~\ref{sec:xi0}.

Taking $\nu_{\rm m,r}$ crossing $i'$ band at the fitted value of 710~s and
$\nu_{\rm m,f}$ crossing at 2400~s, then we can take the typical evolutions of these
frequencies ($t^{-3/2}$ and $t^{-73/48}$ for FS and RS respectively),
to estimate the ratio of frequencies at the deceleration time. This ratio has
a very weak dependence on $t_{\rm d}$ due to the almost identical temporal evolution
of both $\nu_{\rm m,r}$ and $\nu_{\rm m,f}$ and it is $\nu_{\rm m,r}/\nu_{\rm m,f}\simeq0.16$.
Using the numerical results by \citet{Harrison13} and using the definition of $\xi_0$ one
can express this ratio as a function of $\Gamma_0$ and $R_B$ as
\begin{equation}
\frac{\nu_{\rm m,r}(t_{\rm d})}{\nu_{\rm m,f}(t_{\rm d})} = \Big[\frac{5\times10^{-3}}{\Gamma_0^2} + \Big(\frac{c\,T_{90}}{l(1+z)}\Big)^{3/2}\,\Gamma_0^2\Big]\,R_B^{1/2}\;.
\label{eq:nuratio}
\end{equation}
The second constraint comes from the ratio of the maximum flux densities at deceleration,
$F_{\nu,{\rm max,r}}(t_{\rm d})/F_{\nu,{\rm max,f}}$ which can also be expressed as a function
of $\Gamma_0$ and $R_B$ as
\begin{equation}
\frac{F_{\nu,{\rm max,r}}(t_{\rm d})}{F_{\nu,{\rm max,f}}} =\  
0.27\,\Big(\frac{t_{\rm d}}{710\ \textrm{s}}\Big)^{-1}\ =\ \frac{\Gamma_0 R_B^{1/2}}{1.5 + 5\,\xi_0^{-1.3}}\;,
\label{eq:Fmaxratio}
\end{equation}
where $t_{\rm d}$ is given by eq.~(\ref{eq:td}) and $\xi_0=(l/c\,T_{90}\,(1+z))^{1/2}\,\Gamma_0^{-4/3}$.
Equation~(\ref{eq:Fmaxratio}) is derived from numerical results \citep{Harrison13} and using
$F_{\nu,{\rm max,r}}\propto t^{-1}$ \citep{Zhang03}. From our modelling we used
$F_{\nu,{\rm max,r}}(710\ \textrm{s})=0.19$~mJy and $F_{\nu,{\rm max,f}}=0.7$~mJy.
The solution to both eqs.~(\ref{eq:nuratio}) and (\ref{eq:Fmaxratio})
is given by $\Gamma_0\simeq10^4$ and $R_B=4.5$. In spite of the reasonable value for $R_B$, this
scenario appears to be contrived due to the excessively high value for $\Gamma_0$.
We therefore consider energy injection in addition to the RS and examine the evidence for a
continuously refreshed shock.

\subsubsection{An early continuously refreshed shock}
\label{sec:xi0}
In this scenario, we still assume the early time emission originates in the RS, but the
shallow decay phase is caused by energy injection until it switches off.
The outflow profile is such that the slower moving material carries more energy in the system
continuously re--energizing the ejecta as it is envisaged in the refreshed shock scenario \citep{Sari00}.
We consider that the central engine launches material that has a gradient in velocity.
Here the initial deceleration is similar to an impulsive fireball; however the emission is enhanced
as slower moving material catches up with decelerated material.
This makes the decay of the afterglow component shallower.

From $\beta=1.0$ we fitted early dust--corrected $UBVRi'$ fluxes with the combination of the rising FS
as modelled in Section~\ref{sec:boxfit} (which is negligible at $t<800$~s) and
of a continuously refreshed RS in the frequency range
$\nu_{\rm m,r}<\nu_{\rm opt}<\nu_{\rm c,r}$ \citep{Sari00}. The two power--law indices were set
to $\alpha_{\rm r,1}=(12-6s+12\beta)/(2\,(7+s))$ and $\alpha_{\rm r,2}=(73p+21)/96=2.5$ before and after
the end of the energy injection at $t_{\rm r,b}$, respectively \citep{Kobayashi00a}.
The free parameters adopted for the RS contribution were the normalisation, the energy injection end
time $t_{\rm r,b}$, and the velocity profile index $s$, $M(>\gamma)\propto\gamma^{-s}$.
We derive an energy injection parameter $s=3_{-1.1}^{+1.5}$, with energy
injection ending at the observer time $t_{\rm r,b}=930\pm400$~s (Table~\ref{tab:boxfit}).
The result is shown in Figure~\ref{f:boxfit}, where the refreshed RS (RS+FS total)
contribution is shown with dotted (solid) lines.
In spite of the very few additional degrees of freedom, the result is satisfactory.
The overall quality of the RS+FS model of the entire data
set, excluding the X--ray band, improves to $\chi^2/{\rm dof}=176/128$ (p--value of 
0.3\%; Table~\ref{tab:boxfit}), which is still poor, but the overall behaviour
displayed by the data is modelled remarkably well.
The result shows a negligible dependence on the value of the decay index $\alpha_{\rm r,2}$,
because at $t>800$~s the FS component dominates over the RS.
This requires that the amount of energy injected is in the range $2$--$11$~$E_{\gamma,{\rm iso}}$ and directly affects the value inferred for the radiative efficiency $\eta_\gamma$,
which now lies in the range $0.6$--$0.8$, i.e. higher than estimated in eq.~(\ref{eq:eff}).

A final cross--check of this scenario is whether high--latitude emission is affected by energy injection.
Although energy injection switches off at the observer time $t_{\rm r,b}$~s the
high--latitude equivalent lab time emission could, in principle, observe energy injection at later observer
times, thus affecting the FS modelling. However, the high--latitude component decays
more steeply than the line--of--sight component, whose decay index is $\alpha_{\rm r,2}=2.5$,
making it essentially unobservable.

Therefore, in summary, we favour the continuously refreshed reverse shock because, unlike the 
simple reverse shock scenario,  extreme values for $\Gamma_0$ are not required.

\onecolumn
\begin{deluxetable}{ccccccccccccc}
\tablecolumns{13}
\tabletypesize{\scriptsize}
\rotate
\tablecaption{Best--fitting physical parameters obtained from modelling the multi--afterglow
data with the {\sc boxfit} code \citep{Vaneerten12b} combined with an early--time additional
component. Frozen values are in square brackets.
\label{tab:boxfit}}
\tablewidth{0pt}
\tablehead{
\colhead{Dataset\tablenotemark{(a)}} & \colhead{$F_{\rm r,m,i}$\tablenotemark{(b)}} &
\colhead{$t_{\rm r,b}$} & \colhead{$s$} & \colhead{$E_{\rm iso}$} & \colhead{$\theta_{0}$} &
\colhead{$\theta_{\rm obs}/\theta_0$}  & \colhead{$n$}   & \colhead{$p$} &   \colhead{$\epsilon_B$}
& \colhead{$\epsilon_e$} & \colhead{$\xi_N$} & \colhead{$\chi^2/$dof}\\
&   \colhead{(mJy)} & \colhead{(s)} &  & \colhead{($10^{53}$erg)} &  \colhead{($^\circ$)}  & 
 & \colhead{(cm$^{-3}$)} &    &  \colhead{$(10^{-4})$}  &  \colhead{$(10^{-2})$}    &     &   \\ }
\startdata
(1) & -- & -- &  -- & $1.8$      & 26           & $0.96$               & 86        & $3.6$ & $4.7$ & $8.3$ & $[1.0]$ & 451/162\\
(2) & -- & --  &  -- & $1.9_{-0.1}^{+0.7}$     & $23.1_{-4.1}^{+0.8}$  &  $0.93\pm0.01$                & $240_{-90}^{+10}$   & $3.8\pm0.1$ & $2.4_{-0.3}^{+0.1}$ & $9.3_{-3.4}^{+0.5}$ & $0.99$ & 173/122\\
(3)  & $0.37_{-0.14}^{+0.22}$$^{\mathrm{(c)}}$ & $930\pm400^{\mathrm{(c)}}$  & $3.0_{-1.1}^{+1.5}$$^{\mathrm{(c)}}$ & $1.9_{-0.1}^{+0.7}$   & $23.1_{-4.1}^{+0.8}$   & $0.93\pm0.01$               & $240_{-90}^{+10}$       & $3.8\pm0.1$ & $2.4_{-0.3}^{+0.1}$ & $9.3_{-3.4}^{+0.5}$ & $0.99$ & 176/128\\
(3)  & $0.36\pm0.06^{\mathrm{(d)}}$ & $710\pm200^{\mathrm{(d)}}$  & -- & $1.9_{-0.1}^{+0.7}$     & $23.1_{-4.1}^{+0.8}$   & $0.93\pm0.01$  & $240_{-90}^{+10}$  & $3.8\pm0.1$ & $2.4_{-0.3}^{+0.1}$ & $9.3_{-3.4}^{+0.5}$ & $0.99$ & 178/129\\
\enddata
\tablenotetext{(a)}{(1) radio to X, $t>800$~s; (2) radio to UV, $t>800$~s; (3) radio to UV, all.}
\tablenotetext{(b)}{The normalisation is the flux density at the reference time $t_{\rm ref}=100$~s.}
\tablenotetext{(c)}{An additional refreshed RS component was adopted, where the $i'$--band normalisation $F_{\rm r,m,i}$
and end time of energy injection $t_{\rm r,b}$ only were left free to vary. We assumed slow cooling for the RS, with
$\nu_{\rm m,r}<\nu_{\rm opt}<\nu_{\rm c,r}$. $M(>\gamma)\propto\gamma^{-s}$ is the ejected mass moving with Lorentz factors
greater than $\gamma$ \citep{Sari00}.}
\tablenotetext{(d)}{An additional RS component was adopted, where the normalisation $F_{\rm r,m,i}$ and crossing
time of $\nu_{\rm m,r}$ through the $i'$-band only were left free to vary. We assumed slow cooling for the RS.}
\end{deluxetable}
\twocolumn
%

\section{Discussion}
\label{sec:disc}
The most notable and best observed feature of GRB\,120404A is the optical
rebrightening peaking about 40~minutes after the burst, preceded by a nearly
constant flux phase, which appears to be a separate component. While the optical peak
is observed in a number of well sampled early afterglows, it is generally
interpreted as either i) the afterglow onset which marks the deceleration
of the ultra--relativistic ejecta by the circumburst environment or
ii) the FS radiation coming from a jet as seen from an observer outside
the jet cone, i.e. when the viewing angle $\theta_{\rm obs}$ and the
jet half--opening angle $\theta_0$ are such that $\theta_{\rm obs}>\theta_0$.
In the latter case the peak in the light curve
corresponds to the time at which the beaming cone widens enough to become
comparable with the angle from the outer edge of the jet, i.e. when
$1/\Gamma\sim (\theta_{\rm obs}-\theta_0)$ (e.g., \citealt{Granot02, Margutti10}).
In the former case, the peak time is often used to estimate the initial bulk Lorentz
factor at deceleration in the thin shell regime \citep{Sari97}, which is
approximately half its value in the coasting phase preceding the deceleration
(e.g., \citealt{Molinari07,Melandri10,Liang13,Panaitescu13}).

A growing sample of GRBs with exquisite broadband monitoring of the transition
from the end of the prompt emission to the afterglow onset is seen to require
the combination of distinct components to explain all the observations.
In some cases, a double--jet configuration seems to work fairly well (e.g.,
\citealt{Berger03,Huang04,Racusin08,DePasquale09,DePasquale11,Filgas11b,Holland12}).
In other cases, the presence of multiple peaks is explained through the interplay
between RS and FS (e.g., \citealt{Zheng12,Virgili13}, De~Pasquale et al. in prep.),
as expected for given combinations of values for the microphysics parameters and
magnetisation content of the fireball \citep{Kobayashi00a,Zhang03,ZK05}.

Another possibility often discussed is that of energy injection episodes which
keep refreshing the FS, whose complex behaviour would track the history
of how energy is transferred to the FS as a function of time (e.g.,
\citealt{Rossi11,Cucchiara11}). It is not uncommon that some of the best sampled
multi--band afterglows require some combination of these mechanisms
\citep{Greiner13,Virgili13}.

As for GRB\,120404A, while the prompt emission lasts about 50~s, the optical
flux nearly constant with time preceding the rise is detected from $\sim200$
to $\sim800$~s. An internal shock dissipation origin for this early optical emission is
disfavoured because, in contrast to some GRBs with contemporaneous optical and
$\gamma$-ray emission \citep{Kopac13}, no residual $\gamma$-ray activity
is detected beyond the first minute in GRB\,120404A. 
Furthermore, the lack of temporal variability of the early optical
emission argues against an internal shock dissipation origin
(e.g., \citealt{Nardini11}).
An external origin automatically rules out the interpretation of the optical
peak as due to the fireball deceleration.

Unlike the cases above which invoke a hydrodynamical origin for the peak,
an alternative interpretation is that connected with the passage of the
peak synchrotron frequency through the observed bands, which is chromatic
\citep{Sari98}. Although observational evidence for this was reported only for
a few cases, this might be more common than what has currently been found,
simply because many data sets lack well--sampled, simultaneous multi--colour
coverage (\citealt{Oates11,Zheng12}, de~Pasquale et al. in prep.).

Analogous considerations apply to the difficulty of collecting evidence
for a jet break in the afterglow light curves of many GRBs \citep{Racusin09},
whose signature can be more elusive than a clear--cut achromatic break, especially
when different effects come into play simultaneously.

The high quality of the broadband observations of GRB\,120404A show the power
of more comprehensive datasets for severely constraining the energetics, geometry,
and microphysics parameters of the afterglow emission in conjunction with the realistic
{\sc boxfit} fitting code based on hydrodynamics simulations \citep{Vaneerten12b}.
As noted above, this code is applicable when most of the fireball energy has already
been transferred to the ISM, so is used separately to the early time modelling of
the reverse shock emission. In the following sections, we discuss the implications
derived from our modelling: in particular, implications for GRB jet geometries and
the theoretical aspects of the code that could be improved to allow better modelling of the data.

\subsection{The nature of the afterglow peak}
\label{sec:nature_peak}
\begin{figure}
\centering
\includegraphics[width=8.2cm]{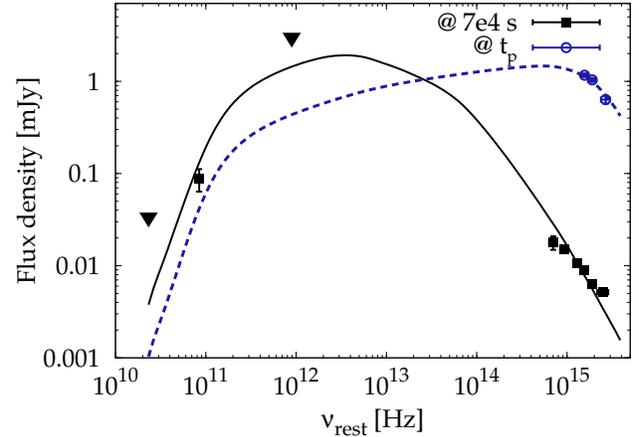}
\caption{Rest--frame SEDs at peak (circles and dashed line) and at $70$~ks (squares
and solid lines) including radio and optical measurements.
Optical points have been corrected for dust extinction using
$A_V=0.22$~mag for a MW profile. Upside--down triangles are 3$\sigma$ upper limits.
The thick solid line shows the best--fitting model based on hydrodynamical
simulations.}
\label{f:radiosed}
\end{figure}

The afterglow peak is the result of the passage of $\nu_{\rm m,f}$ through
the optical bands \citep{Zhang03}, as was the case for other exquisitely sampled GRBs
(e.g., \citealt{Zheng12}). To show this, we obtained two SEDs: one is measured
at the peak time, the other refers to 70~ks after the burst. Figure~\ref{f:radiosed}
displays the two SEDs together with the models corresponding to the set of best--fitting
parameters obtained above. Noteworthy is how our radio measurements are fully
consistent with the broadband evolution and clearly show the self--absorbed
regime of the synchrotron spectrum in the late SED.
%
%

The model predicts a steeper slope than that exhibited by the optical data points.
This is connected with the global cooling approximation issue: due to this,
the model in Figure~\ref{f:radiosed} places the cooling frequency $\nu_{\rm c,f}$
below the optical points, while a correct treatment of the local cooling
would place it well above (see Fig.~4 of \citealt{Vaneerten10a}),
thus explaining both the common spectral regime between optical and X--rays
as well as the normalisation of the observed X--ray flux.
In this case, the need for matching the radio and the optical fluxes with a more
plausible optical slope than the best--fitting model shown in Fig.~\ref{f:radiosed}
would require the FS peak flux density in frequency, $F_{\nu,{\rm max,f}}$, to decrease
with time. However, this clashes with the $F_{\nu,{\rm max,f}}\propto t^0$ evolution
expected in the homogeneous medium scenario assumed by the {\sc boxfit} code.
For a wind density profile, it is $F_{\nu,{\rm max,f}}\propto t^{-1/2}$ \citep{Chevalier99}.
In Section~\ref{sec:broadband} we argued that a wind--like density profile is not ruled
out from the expected closure relation at late times.
This suggests that a local cooling treatment combined with the possibility of wind--like
environments could help to improve the modelling capabilities of the {\sc boxfit} code.
A repeat run {\em without} cooling modelled the optical slope slightly better, but did
not improve the quality of the overall fit and caused a significantly worse fit to the radio data.
This therefore further confirms the need for a code development to include more realistic
cooling and density profiles.

\subsection{An edge--on view of a wide jet}
\begin{figure}
\centering
\includegraphics[width=8.5cm]{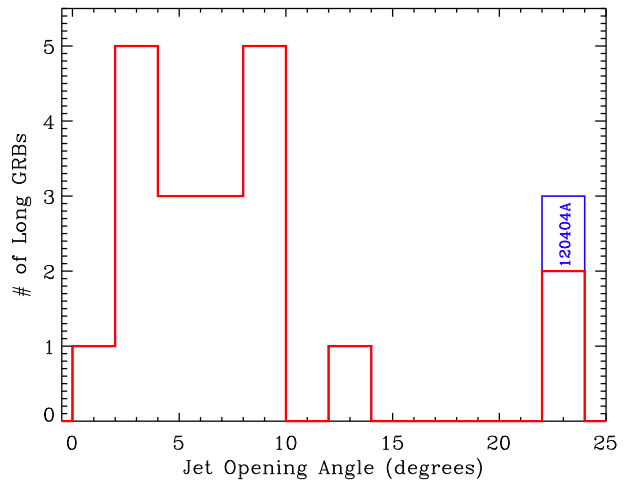}
\caption{Jet half--opening angle distribution for a number of {\em Swift}
long GRBs. GRB\,120404A lies in the wide--angle tail.}
\label{f:jetdist}
\end{figure}

Excluding the X--ray data, the best--fitting parameters do not change in essence and
confirm the basic picture of a relatively wide jet viewed from a direction close
to the edge but still inside the jet cone. Although the true $\theta_0$ distribution
is difficult to derive from observations because of the numerous selection effects
and observational biases \citep{Bloom03,Lu12}, past data suggest the existence
of comparably wide jets \citep{Bloom03,Fong12}, as clearly shown in
Fig.~\ref{f:jetdist} which displays the $\theta_0$ distribution for a number of
{\em Swift} long GRBs.
The FS microphysics parameters are within the range of typical values estimated for
other GRBs \citep{Panaitescu02}, apart from the high value of $p$, which is likely
to be connected with the afore--mentioned global cooling issue. The ISM particle
density $n$ is high, but still within the high tail of the distribution.

%
%

%
\begin{figure}
\centering
\includegraphics[width=8.5cm]{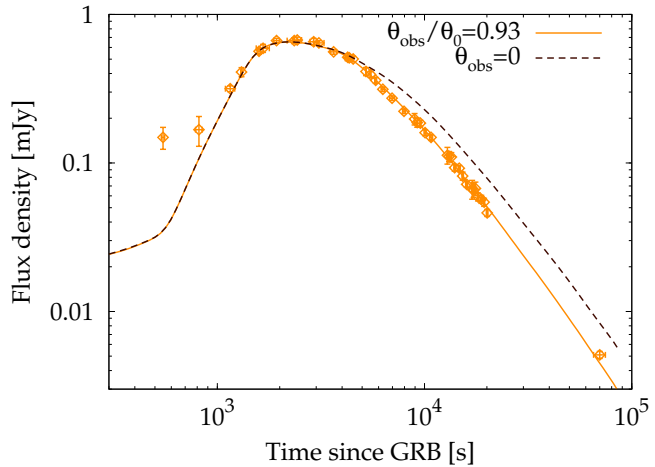}
\caption{Light curve in the $i'$--band. The solid line shows the best--fitting
solution obtained for a jet opening angle $\theta_0=23^\circ$ and viewing
angle $\theta_{\rm obs}=21^\circ$. The dashed
line shows what an on--axis observer would have observed for the same GRB.}
\label{f:on_vs_offaxis}
\end{figure}
%

The quality of the data set, combined with the capability of the fitting code,
allowed us to constrain both $\theta_0$ and $\theta_{\rm obs}$, as shown by
Figure~\ref{f:on_vs_offaxis}, which compares the observed data and model with
what an on--axis observer would have observed.
The accurately modelled shape of the multi--band light
curves is sensitive to the missing flux from the jet edge closer to the observer
sightline, when $1/\Gamma\sim (\theta_0-\theta_{\rm obs})$. The time at which
this is observed can be clearly estimated from Fig.~\ref{f:on_vs_offaxis} around
$t_{j,1}\simeq 5$~ks, and corresponds to
\begin{equation}
\theta_0-\theta_{\rm obs} = 2.7^\circ\ \Big(\frac{t_{j,1}\,\zeta}{5\ \textrm{ks}}\Big)^{3/8}\ 
\Big(\frac{n_0}{244}\Big)^{1/8}\ \Big(\frac{E_{{\rm iso},53}}{1.9}\Big)^{-1/8}\;,
\label{eq:tj1}
\end{equation}
as expected from the corresponding best--fitting values ($\zeta=3.876/(1+z)$).
One should expect to also see the final and steep drop in the decay slope
associated with the further jet edge, i.e. when it is $\Gamma\sim 1/(\theta_0+\theta_{\rm obs})$.
This is to be expected at the time $t_{j,2}$
\begin{equation}
t_{j,2} = \Big(\frac{\theta_0+\theta_{\rm obs}}{\theta_0-\theta_{\rm obs}}\Big)^{8/3}\ t_{j,1}\simeq
100\ \textrm{days}\;,
\label{eq:tj2}
\end{equation}
which is far beyond the coverage of our data set.

Another interesting result from the afterglow modelling is the possibility to constrain
the radiative efficiency $\eta_\gamma$ of the prompt emission \citep{Zhang07},
\begin{equation}
\eta_\gamma\ =\ \frac{E_{\gamma,{\rm iso}}}{E_{\rm iso} + E_{\gamma,{\rm iso}}}\ =\ 0.3 \pm 0.1\;,
\label{eq:eff}
\end{equation}
which is consistent with typical expectations from internal shocks
\citep{Beloborodov00,Guetta01,Kobayashi01} as well as with values measured for other GRBs \citep{Zhang07}.
However, the value of $0.3$ for $\eta_\gamma$ becomes a lower limit if the early optical emission
is due to prolonged internal activity, such as the case considered in Section~\ref{sec:xi0}.
It is also possible to calculate the total kinetic energy corrected for collimation,
\begin{equation}
E_{\rm K}\ =\ E_{\rm iso}\,(1-\cos{\theta_0})\ =\ 1.5\times10^{52}\ \textrm{erg}\;.
\label{eq:Ek}
\end{equation}
In addition to the global cooling approximation issue, other limitations of the {\sc boxfit}
code concern the jet angular structure, assumed to be homogeneous with sharp boundaries,
as well as the assumption of a homogeneous instead of a wind--like density profile
of the surrounding medium, as argued in Section~\ref{sec:nature_peak}.
Adopting more realistic jet structures can possibly lead to further improvements in the data
modelling (e.g., \citealt{Rossi02,Zhang02,Zhang04,Guetta05,Granot05,Panaitescu08}).

\section{Conclusions}
\label{sec:conc}
We presented the first high--quality broadband data set of a GRB fitted
with a realistic code developed and made available by \citet{Vaneerten12b}.
This code was built upon hydrodynamical simulations and not
merely on analytical approximations, to model the afterglow evolution
from radio to high--energies.
In particular, we found that synchrotron radiation expected from
the forward shock propagating through a constant medium within the shape
of a homogeneous jet with opening angle $\theta_0=23.1_{-4.1}^{+0.8}$~degrees
viewed almost edge--on, $\theta_{\rm obs}=(0.93\pm0.01)\,\theta_0$ can
reproduce very accurately the well sampled multi--band light curves of GRB\,120404A.

We constrained the microphysics of the relativistic shock,
which gives rise to the forward shock emission and a--posteriori highlights
the importance of adopting a local cooling treatment in place of the commonly
adopted global cooling in modelling the flux above the cooling break.
These results unambiguously suggest that future theoretical developments and
refinements of models like the one we adopted here should provide a more realistic
description of the local cooling and its impact at high energies, and
should also include wind--like density profiles in addition to the already
treated homogeneous case.

The optical peak observed a few thousands seconds after the burst, which
is a common property among many GRB early afterglows, shows evidence for
a chromatic character and is satisfactorily explained by the crossing
of the FS synchrotron peak frequency $\nu_{\rm m,f}$ through the observed
bands. This exclude the interpretation of the peak as the time of fireball deceleration,
which appears to be contrived due to the initial constant optical flux density,
\footnote{To be explained in terms of fireball deceleration, it would require a
non--uniform shell with the innermost part carrying a significant fraction of
the shell energy, so that the interstellar medium would receive a major impulsive
energy injection at the end of the reverse shock crossing.} and highlight the need for
caution in automatically interpreting all optical peaks as deceleration signatures.

We instead successfully modelled the same early optical emission in terms of
a reverse shock which is continuously refreshed by a velocity distribution
of the ejecta for about $10^3$~s after the burst.
The fireball deceleration occurs in the intermediate/thick
shell regime and highlights the importance of  correct treatment to evaluate
the relative strength between forward and reverse shock emission to constrain
the fireball magnetisation and the initial bulk Lorentz factor \citep{Harrison13}.

The total released isotropic--equivalent and collimation--corrected energy,
$\sim2\times$ and $\sim0.1\times$ $10^{53}$~erg respectively, allowed
us to directly estimate the radiative efficiency of the prompt emission,
which is found to be either $\eta_\gamma=0.7\pm0.15$ or $\eta_\gamma=0.3\pm0.1$,
depending on whether the early optical emission is the result or not of
prolonged energy injection into the fireball.

Our capability of constraining the jet geometry provides new insight into the long-standing  difficulty of measuring clear--cut jet breaks for GRBs  in the {\it Swift} era and emphasises the role
played by an off--axis angle when this is comparable to the jet opening angle.
Moreover, our results add strong support for the growing evidence that comparably wide jets as viewed
from comparably large off--axis directions are probably more common than previously
inferred from simple analytical descriptions \citep{Ryan13}.

Overall, the entire data set is well explained with a combination of 
reverse and forward shocks in a relatively wide homogeneous jet viewed nearly edge--on
plus energy injection.
This seems to be a natural choice in other GRBs similar to GRB\,120404A, in which the prompt
emission is characterised by a very simple FRED.

\section*{Acknowledgments}

C.G. acknowledges ASI for financial support (ASI-INAF contract
I/088/06/0) and University of Ferrara for use of the local HPC facility
co-funded by the ``Large-Scale Facilities 2010'' project (grant 7746/2011).
C.G. thanks Hendrik van Eerten for useful discussions.
C.G.M. thanks the Royal Society, Wolfson Foundation and Science and Technology Facilities Council for support.
E.~B. and B.~A.~Z. were supported in part by the National Science Foundation under
Grant AST-1107973. B.~A.~Z. is supported by an NSF Astronomy and Astrophysics
Postdoctoral Fellowship under award AST-1302954.
A.G. acknowledges founding from the Slovenian Research Agency and from the
Centre of Excellence for Space Sciences and Technologies SPACE-SI, an operation
partly financed by the European Union, the European Regional
Development Fund, and the Republic of Slovenia, Ministry of Education, Science and Sport.
Development of the Boxfit code was supported in part by NASA through grant
NNX10AF62G issued through the Astrophysics Theory Program and by the NSF through
grant AST-1009863. 
The NRAO is a facility of the NSF operated under cooperative agreement by AUI.
Proposal VLA~12A-394 is acknowledged.
The SMA is a joint project between SAO and the ASIAA and is funded by the
Smithsonian Institution and the Academia Sinica.
Part of the funding for GROND (both hardware as well as personnel) was generously
granted from the Leibniz-Prize to Prof. G. Hasinger (DFG grant HA 1850/28-1).
DARK is funded by the DNRF.
We thank the anonymous referee for helpful comments that improved the paper.


\twocolumn
\begin{deluxetable}{rcrclcrcrclc}
\tablecolumns{12}
\tabletypesize{\scriptsize}
\tablecaption{Photometric data set of the NIR/optical afterglow of 
GRB\,120404A.}
\label{tab:phot}
\tablewidth{0pt}
\tablehead{
\colhead{Time\tablenotemark{a}} & \colhead{Telescope} & \colhead{Exp.} &
\colhead{Filter} & \colhead{Magnitude\tablenotemark{b}} & \colhead{Flux\tablenotemark{c}} &
\colhead{Time\tablenotemark{a}} & \colhead{Telescope} & \colhead{Exp.} &
\colhead{Filter} & \colhead{Magnitude\tablenotemark{b}} & \colhead{Flux\tablenotemark{c}}\\
\colhead{(s)} & & \colhead{(s)} & & & \colhead{($\mu$Jy)} &
\colhead{(s)} & & \colhead{(s)} & & & \colhead{($\mu$Jy)}}
\startdata
     70023 & GROND &   4592 & $K$\tablenotemark{d}  & $>$ $20.5$      & $<$ $23.3$          &   9134 & FTS &       30 & $R$  & $18.36\pm 0.08$ & $  157.6\pm   11.2$\\
     70023 & GROND &   4592 & $H$\tablenotemark{d}  & $21.1 \pm 0.2$  & $   13.5\pm    2.3$ &   9428 & FTS &       60 & $R$  & $18.34\pm 0.06$ & $  160.5\pm    8.6$\\
     70023 & GROND &   4592 & $J$\tablenotemark{d}  & $21.4 \pm 0.1$  & $   10.4\pm    0.9$ &   9873 & FTS &      120 & $R$  & $18.46\pm 0.05$ & $  143.7\pm    6.5$\\
     70023 & GROND &   4592 & $z'$ & $22.09\pm 0.06$ & $    5.7\pm    3.1$ &  10512 & FTS &      180 & $R$  & $18.61\pm 0.05$ & $  125.2\pm    5.6$\\
     25370 & ARIES &    900 & $I$  & $19.71\pm 0.17$ & $   36.4\pm    5.3$ &  12619 & FTS &       30 & $R$  & $18.88\pm 0.11$ & $   97.6\pm    9.4$\\
       547 & FTN &       10 & $i'$ & $18.58\pm 0.20$ & $  148.6\pm   25.0$ &  13072 & FTS &       30 & $R$  & $18.96\pm 0.11$ & $   90.7\pm    8.7$\\
       816 & FTN &       30 & $i'$ & $18.45\pm 0.28$ & $  167.5\pm   38.1$ &  13352 & FTS &       60 & $R$  & $18.89\pm 0.09$ & $   96.7\pm    7.7$\\
      1154 & FTN &       60 & $i'$ & $17.76\pm 0.05$ & $  316.2\pm   14.2$ &  13780 & FTS &      120 & $R$  & $19.04\pm 0.07$ & $   84.2\pm    5.3$\\
      1308 & FTS &       10 & $i'$ & $17.48\pm 0.08$ & $  409.3\pm   29.1$ &  14469 & FTS &      180 & $R$  & $19.09\pm 0.05$ & $   80.4\pm    3.6$\\
      1589 & FTS &       30 & $i'$ & $17.13\pm 0.05$ & $  565.0\pm   25.4$ &  15092 & FTS &      120 & $R$  & $19.23\pm 0.08$ & $   70.7\pm    5.0$\\
      1662 & FTN &      120 & $i'$ & $17.08\pm 0.07$ & $  591.6\pm   36.9$ &  15724 & FTS &      180 & $R$  & $19.24\pm 0.07$ & $   70.1\pm    4.4$\\
      1932 & FTS &       60 & $i'$ & $16.95\pm 0.04$ & $  666.8\pm   24.1$ &  16689 & FTS &       30 & $R$  & $19.50\pm 0.15$ & $   55.1\pm    7.1$\\
      2352 & FTN &      180 & $i'$ & $16.95\pm 0.05$ & $  666.8\pm   30.0$ &  17177 & FTS &       30 & $R$  & $19.38\pm 0.15$ & $   61.6\pm    7.9$\\
      2432 & FTS &      120 & $i'$ & $16.94\pm 0.05$ & $  673.0\pm   30.3$ &  17480 & FTS &       60 & $R$  & $19.66\pm 0.19$ & $   47.6\pm    7.6$\\
      2918 & FTN &      120 & $i'$ & $16.97\pm 0.06$ & $  654.7\pm   35.2$ &  17919 & FTS &      120 & $R$  & $19.54\pm 0.08$ & $   53.1\pm    3.8$\\
      3093 & FTS &      180 & $i'$ & $16.99\pm 0.04$ & $  642.7\pm   23.2$ &  18549 & FTS &      180 & $R$  & $19.66\pm 0.07$ & $   47.6\pm    3.0$\\
      3643 & FTS &      120 & $i'$ & $17.14\pm 0.04$ & $  559.8\pm   20.2$ &  19181 & FTS &      120 & $R$  & $19.70\pm 0.07$ & $   45.9\pm    2.9$\\
      4256 & FTN &       10 & $i'$ & $17.23\pm 0.06$ & $  515.3\pm   27.7$ &  19824 & FTS &      180 & $R$  & $19.76\pm 0.05$ & $   43.4\pm    2.0$\\
      4331 & FTS &      180 & $i'$ & $17.24\pm 0.03$ & $  510.5\pm   13.9$ &  24668 & ARIES &   1200 & $R$  & $20.37\pm 0.15$ & $   24.7\pm    3.2$\\
      4526 & FTN &       30 & $i'$ & $17.26\pm 0.05$ & $  501.2\pm   22.6$ &  57288 & VLT/XS &    30 & $R$  & $21.4\pm  0.3$  & $    9.6\pm    2.3$\\
      5219 & FTS &       10 & $i'$ & $17.47\pm 0.07$ & $  413.1\pm   25.8$ &  70023 & GROND &   4592 & $r'$ & $22.61\pm 0.06$ & $    3.8\pm    0.2$\\
      5484 & FTS &       30 & $i'$ & $17.52\pm 0.05$ & $  394.5\pm   17.8$ &    471 & FTN &       10 & $V$  & $18.70\pm 0.30$ & $  139.9\pm   33.8$\\
      5810 & FTS &       60 & $i'$ & $17.62\pm 0.05$ & $  359.8\pm   16.2$ &    878 & UVOT &     400 & $V$  & $18.83\pm 0.33$ & $  124.1\pm   32.5$\\
      6294 & FTS &      120 & $i'$ & $17.77\pm 0.04$ & $  313.3\pm   11.3$ &   1237 & FTS &       10 & $V$  & $17.97\pm 0.16$ & $  274.1\pm   37.6$\\
      6970 & FTS &      180 & $i'$ & $17.92\pm 0.04$ & $  272.9\pm    9.9$ &   4191 & FTN &       10 & $V$  & $17.61\pm 0.09$ & $  381.9\pm   30.4$\\
      7956 & FTS &      120 & $i'$ & $18.14\pm 0.04$ & $  222.9\pm    8.1$ &   4832 & UVOT &     200 & $V$  & $17.74\pm 0.10$ & $  338.8\pm   29.8$\\
      8957 & FTS &       10 & $i'$ & $18.27\pm 0.10$ & $  197.7\pm   17.4$ &   5144 & FTS &       10 & $V$  & $17.74\pm 0.12$ & $  338.8\pm   35.4$\\
      9219 & FTS &       30 & $i'$ & $18.33\pm 0.08$ & $  187.1\pm   13.3$ &   8892 & FTS &       10 & $V$  & $18.48\pm 0.21$ & $  171.4\pm   30.1$\\
      9552 & FTS &       60 & $i'$ & $18.34\pm 0.05$ & $  185.4\pm    8.3$ &   8941 & UVOT &    5545 & $V$  & $18.90\pm 0.13$ & $  116.4\pm   13.1$\\
     10061 & FTS &      120 & $i'$ & $18.50\pm 0.05$ & $  160.0\pm    7.2$ &  11992 & UVOT &     550 & $V$  & $19.42\pm 0.18$ & $   72.1\pm   11.0$\\
     10754 & FTS &      180 & $i'$ & $18.58\pm 0.05$ & $  148.6\pm    6.7$ &  14858 & FTS &       20 & $V$  & $19.57\pm 0.35$ & $   62.8\pm   17.3$\\
     12889 & FTS &       10 & $i'$ & $18.88\pm 0.15$ & $  112.7\pm   14.5$ &  70023 & GROND &   4592 & $g'$ & $23.1 \pm 0.1$  & $    2.5\pm    0.2$\\
     13151 & FTS &       30 & $i'$ & $18.90\pm 0.09$ & $  110.7\pm    8.8$ &    413 & FTN &       10 & $B$  & $19.43\pm 0.27$ & $   87.8\pm   19.3$\\
     13473 & FTS &       60 & $i'$ & $18.91\pm 0.08$ & $  109.7\pm    7.8$ &    649 & FTN &       30 & $B$  & $19.38\pm 0.24$ & $   91.9\pm   18.2$\\
     13967 & FTS &      120 & $i'$ & $19.09\pm 0.06$ & $   92.9\pm    5.0$ &    891 & UVOT &     572 & $B$  & $19.05\pm 0.20$ & $  124.6\pm   21.0$\\
     14710 & FTS &      180 & $i'$ & $19.10\pm 0.05$ & $   92.1\pm    4.1$ &    922 & FTN &       60 & $B$  & $19.16\pm 0.12$ & $  112.6\pm   11.8$\\
     15276 & FTS &      120 & $i'$ & $19.23\pm 0.07$ & $   81.7\pm    5.1$ &   1340 & UVOT &      20 & $B$  & $18.32\pm 0.25$ & $  244.0\pm   50.2$\\
     15967 & FTS &      180 & $i'$ & $19.37\pm 0.05$ & $   71.8\pm    3.2$ &   1402 & FTS &       30 & $B$  & $18.21\pm 0.30$ & $  270.0\pm   65.2$\\
     16971 & FTS &       10 & $i'$ & $19.45\pm 0.17$ & $   66.7\pm    9.7$ &   1691 & FTS &       60 & $B$  & $18.16\pm 0.13$ & $  282.8\pm   31.9$\\
     17256 & FTS &       30 & $i'$ & $19.46\pm 0.13$ & $   66.1\pm    7.5$ &   1874 & FTN &      180 & $B$  & $18.19\pm 0.39$ & $  275.0\pm   83.0$\\
     17598 & FTS &       60 & $i'$ & $19.44\pm 0.12$ & $   67.3\pm    7.0$ &   2080 & FTS &      120 & $B$  & $17.95\pm 0.08$ & $  343.1\pm   24.4$\\
     18108 & FTS &      120 & $i'$ & $19.58\pm 0.07$ & $   59.2\pm    3.7$ &   2567 & FTN &      120 & $B$  & $18.05\pm 0.11$ & $  312.9\pm   30.1$\\
     18796 & FTS &      180 & $i'$ & $19.62\pm 0.07$ & $   57.0\pm    3.6$ &   2628 & FTS &      180 & $B$  & $18.00\pm 0.07$ & $  327.6\pm   20.5$\\
     19379 & FTS &      120 & $i'$ & $19.67\pm 0.07$ & $   54.5\pm    3.4$ &   3297 & FTS &      120 & $B$  & $18.16\pm 0.10$ & $  282.8\pm   24.9$\\
     20053 & FTS &      180 & $i'$ & $19.85\pm 0.06$ & $   46.1\pm    2.5$ &   3858 & FTS &      180 & $B$  & $18.28\pm 0.08$ & $  253.2\pm   18.0$\\
     70023 & GROND &   4592 & $i'$ & $22.24\pm 0.06$ & $    5.1\pm    2.7$ &   4131 & FTN &       10 & $B$  & $18.38\pm 0.11$ & $  230.9\pm   22.2$\\
       246 & FTN &       10 & $R$  & $18.36\pm 0.13$ & $  157.6\pm   17.8$ &   4354 & FTN &       30 & $B$  & $18.52\pm 0.08$ & $  203.0\pm   14.4$\\
       283 & FTN &       10 & $R$  & $18.40\pm 0.16$ & $  151.9\pm   20.8$ &   5200 & FTS &       40 & $B$  & $18.60\pm 0.27$ & $  188.5\pm   41.5$\\
       323 & FTN &       10 & $R$  & $18.25\pm 0.21$ & $  174.4\pm   30.7$ &   5582 & FTS &       60 & $B$  & $18.87\pm 0.18$ & $  147.0\pm   22.5$\\
       730 & FTN &       30 & $R$  & $18.39\pm 0.23$ & $  153.3\pm   29.3$ &   5652 & UVOT &     200 & $B$  & $18.65\pm 0.10$ & $  180.1\pm   15.8$\\
      1001 & FTS &       10 & $R$  & $18.03\pm 0.13$ & $  213.5\pm   24.1$ &   5954 & FTS &      120 & $B$  & $18.78\pm 0.11$ & $  159.7\pm   15.4$\\
      1036 & FTN &       60 & $R$  & $17.97\pm 0.07$ & $  225.7\pm   14.1$ &   6509 & FTS &      180 & $B$  & $18.98\pm 0.11$ & $  132.9\pm   12.8$\\
      1041 & FTS &       10 & $R$  & $18.01\pm 0.13$ & $  217.5\pm   24.5$ &   7088 & UVOT &     200 & $B$  & $18.86\pm 0.12$ & $  148.4\pm   15.5$\\
      1082 & FTS &       10 & $R$  & $17.92\pm 0.15$ & $  236.3\pm   30.5$ &   7605 & FTS &      120 & $B$  & $19.21\pm 0.15$ & $  107.5\pm   13.9$\\
      1479 & FTN &      120 & $R$  & $17.16\pm 0.09$ & $  475.8\pm   37.9$ &   8173 & FTS &      180 & $B$  & $19.31\pm 0.12$ & $   98.0\pm   10.3$\\
      1500 & FTS &       30 & $R$  & $17.09\pm 0.05$ & $  507.5\pm   22.8$ &   9088 & FTS &      100 & $B$  & $19.52\pm 0.20$ & $   80.8\pm   13.6$\\
      1806 & FTS &       60 & $R$  & $16.98\pm 0.06$ & $  561.7\pm   30.2$ &   9699 & FTS &      120 & $B$  & $19.28\pm 0.17$ & $  100.8\pm   14.6$\\
      2106 & FTN &      180 & $R$  & $16.90\pm 0.03$ & $  604.6\pm   16.5$ &  10266 & FTS &      180 & $B$  & $19.76\pm 0.16$ & $   64.8\pm    8.9$\\
      2249 & FTS &      120 & $R$  & $16.87\pm 0.03$ & $  621.5\pm   16.9$ &  13208 & FTS &      220 & $B$  & $20.24\pm 0.20$ & $   41.6\pm    7.0$\\
      2735 & FTN &      120 & $R$  & $16.92\pm 0.08$ & $  593.6\pm   42.2$ &  14166 & FTS &      180 & $B$  & $20.39\pm 0.22$ & $   36.3\pm    6.7$\\
      2857 & FTS &      180 & $R$  & $16.89\pm 0.03$ & $  610.2\pm   16.6$ &  15611 & FTS &      540 & $B$  & $20.75\pm 0.20$ & $   26.0\pm    4.4$\\
      3470 & FTS &      120 & $R$  & $17.02\pm 0.03$ & $  541.3\pm   14.8$ &  17113 & UVOT &     907 & $B$  & $20.51\pm 0.16$ & $   32.5\pm    4.4$\\
      3989 & FTN &       30 & $R$  & $17.15\pm 0.04$ & $  480.3\pm   17.4$ &  18243 & FTS &      700 & $B$  & $20.95\pm 0.20$ & $   21.6\pm    3.6$\\
      4090 & FTS &      180 & $R$  & $17.16\pm 0.03$ & $  475.8\pm   13.0$ &    474 & UVOT &     246 & $U$  & $20.06\pm 0.26$ & $   17.6\pm    4.2$\\
      4436 & FTN &       30 & $R$  & $17.26\pm 0.04$ & $  434.0\pm   15.7$ &   1039 & UVOT &      58 & $U$  & $19.35\pm 0.36$ & $   33.7\pm   11.3$\\
      4950 & FTS &       30 & $R$  & $17.38\pm 0.05$ & $  388.6\pm   17.5$ &   5447 & UVOT &     197 & $U$  & $18.97\pm 0.15$ & $   47.9\pm    6.4$\\
      5398 & FTS &       30 & $R$  & $17.52\pm 0.05$ & $  341.6\pm   15.4$ &   6883 & UVOT &     197 & $U$  & $19.49\pm 0.20$ & $   29.6\pm    5.6$\\
      5698 & FTS &       60 & $R$  & $17.60\pm 0.05$ & $  317.3\pm   14.3$ &  28449 & UVOT &     396 & $U$  & $>$ $21.0$      & $<$ $7.6$\\
      6121 & FTS &      120 & $R$  & $17.69\pm 0.04$ & $  292.1\pm   10.6$ & 104796 & UVOT &    1678 & $U$  & $>$ $21.8$      & $<$ $3.4$\\
      6738 & FTS &      180 & $R$  & $17.84\pm 0.04$ & $  254.4\pm    9.2$ & 173817 & UVOT &    1687 & $U$  & $>$ $21.8$      & $<$ $3.4$\\
      7780 & FTS &      120 & $R$  & $18.06\pm 0.05$ & $  207.7\pm    9.3$ & 256985 & UVOT &    1687 & $U$  & $>$ $21.8$      & $<$ $3.4$\\
      8717 & FTS &       30 & $R$  & $18.24\pm 0.07$ & $  176.0\pm   11.0$ &        &      &         &      &                 &          \\
\enddata
\tablecomments{Uncertainties are 1$\sigma$.}
\tablenotetext{a}{Midpoint time from the GRB onset time.}
\tablenotetext{b}{Corrected for airmass.}
\tablenotetext{c}{Corrected for Galactic extinction.}
\tablenotetext{d}{$AB$ magnitudes.}
\end{deluxetable}


\begin{table*}
\centering
  \caption{Radio data.}
  \label{tab:radio}
  \begin{tabular}{ccccccccc}
\hline
Frequency$^{\mathrm{a}}$ & Start time$^{\mathrm{b}}$ & Int src$^{\mathrm{c}}$ & RMS$^{\mathrm{d}}$ &
Integrated flux$^{\mathrm{e}}$ & Beam size$^{\mathrm{f}}$ & Observatory\\
(GHz) & (days) & (min) & ($\mu$Jy/bm) & ($\mu$Jy) & ($''$) &\\
\hline
$21.8*$ &  $0.75$ &  $17.5$ &  $25.3$ &  $87.6\pm 24.0$ &  $1.1\times 0.85$  & VLA\\
$19.1$  &    -    &    -    &  $31.8$ &  $70.6\pm 30.2$ &  $1.2\times 0.94$  &  -\\
$24.4$  &    -    &    -    &  $39.0$ & $112.7\pm 37.0$ &  $0.97\times 0.77$ &  -\\
\hline
$6.0*$  &  $0.77$ &  $23.3$ &  $11.2$ & $< 33.6$        &  $3.1\times 2.7$   & VLA\\
$4.9$   &    -    &    -    &  $17.2$ & $< 51.6$        &  $3.8\times 3.5$   &  -\\
$7.0$   &    -    &    -    &  $14.0$ & $< 42.0$        &  $2.7\times 2.5$   &  -\\
\hline
$230.5$ &  $0.77$ & $302.5$ &  $\sim 10^3$  & $\la 3\times10^3$ &  $0.5\times 0.4$   & SMA\\
\hline
\end{tabular}
 \begin{list}{}{}
  \item[$^{\mathrm{a}}$Mean frequency of observations. The asterisk indicates redundancy, as it is the
        mean of two sidebands listed below.]
  \item[$^{\mathrm{b}}$Since the GRB.]
  \item[$^{\mathrm{c}}$Integration time on source.]
  \item[$^{\mathrm{d}}$Measured with IMSTAT in AIPS.]
  \item[$^{\mathrm{e}}$Integrated flux using AIPS task JMFIT to fit a Gaussian, fixing size to clean beam.] 
  \item[$^{\mathrm{f}}$Synthesized clean beam size. The source for the K band detection is
        not resolved, so this is just the beam size.]
  \end{list}
\end{table*}

\end{document}